\documentclass[epsfig, usenatbib, usegraphicx]{mn2e}
\usepackage{epsfig}
\usepackage{natbib}
\usepackage{graphicx}
\usepackage{fixltx2e}
\usepackage{multirow}
\usepackage{amsmath}
\usepackage{amssymb}

\title{The evolution of Luminous Red Galaxies in the Sloan Digital Sky Survey 7th data release }
\author[Tojeiro et al.]{Rita Tojeiro\thanks{E-mail: rita.tojeiro@port.ac.uk}$^1$ and 
  Will J. Percival$^1$\\
$^1$Institute of Cosmology and Gravitation, Dennis Sciama Building, University of Portsmouth,
Burnaby Road, Portsmouth, PO1 3FX \\
}

\def\gs{\mathrel{\raise1.16pt\hbox{$>$}\kern-7.0pt %
\lower3.06pt\hbox{{$\scriptstyle \sim$}}}}         %
\def\ls{\mathrel{\raise1.16pt\hbox{$<$}\kern-7.0pt %
\lower3.06pt\hbox{{$\scriptstyle \sim$}}}}         %

\newcommand{\vmax}{$V_{\rm max}$ }
\newcommand{\vmatch}{$V_{\rm match}$ }

\newcommand{\mpcoh}{\,h^{-1}\,{\rm Mpc}}

\begin{document}

\maketitle

\begin{abstract}
  We present a comprehensive study of the evolution of Luminous Red
  Galaxies (LRGs) in the latest and final spectroscopic data release
  of the Sloan Digital Sky Survey. We test the scenario of passive
  evolution of LRGs in $0.15<z<0.5$, by looking at the evolution of
  the number and luminosity density of LRGs, as well as of their
  clustering. A new weighting scheme is introduced that allows us to
  keep a large number of galaxies in our sample and put stringent
  constraints on the growth and merging allowed by the data as a
  function of galaxy luminosity. Introducing additional
  luminosity-dependent weighting for our clustering analysis allows us
  to additionally constrain the nature of the mergers. We find that, in
  the redshift range probed, the population of LRGs grows in
  luminosity by 1.5-6 \% Gyr$^{-1}$ depending on their
  luminosity. This growth is predominantly happening in objects that
  reside in the lowest-mass haloes probed by this study, and cannot be explained by satellite
  accretion into massive LRGs, nor by LRG-LRG merging. We find that
  the evolution of the brightest objects (with a K+e-corrected $M_{r,0.1} \lesssim -22.8$) is consistent with that
  expected from passive evolution.
\end{abstract}

\begin{keywords}
galaxies: evolution - cosmology: observations - surveys
\end{keywords}

\title{LRG evolution in DR7}

\section{Introduction}  \label{sec:intro}

Luminous Red Galaxies (LRGs) are useful probes of large scale
structure. They are bright and have the potential to be easily
separated by colour, allowing a uniform
sample to be selected with which to map large volumes. They are also
strongly biased with respect to the matter distribution, making them
particularly attractive for studies that aim to use Baryon Acoustic
Oscillations (BAO) in the galaxy distribution to study the expansion
of the Universe \citep{SeoEtAl03,BlakeEtAl03,HuHaiman03,Matsubara04}. Because of
this, LRGs are being targeted by the Baryon Oscillation Spectroscopic
Survey (BOSS; \citealt{SchlegelEtAl09}), part of the SDSS-III project,
in order to map the expansion rate out to $z<0.7$.

In addition to allowing measurements of cosmological expansion by
using the BAO signal as a standard ruler, galaxy surveys also
measure the growth of structure through redshift-space distortions
(e.g., \citealt{Kaiser87, DavisPeebles83} and
\citealt{Hamilton98} for a review). Although current BAO and
z-space distortion analyses are generally robust to an evolving
galaxy bias, future experiments will be more sensitive to these
effects (e.g. \citealt{SmithEtAl07}). Observing a single population of 
passively evolving galaxies offers an attractive way of minimising 
bias effects. In particular, for such a sample, the redshift 
evolution of bias and its effects on cosmological probes can be 
more easily modelled. Uniformity in terms of selection is also 
advantageous for cosmological studies of large-scale structure, 
which are often simplified if the relationship between the galaxies 
and the underlying density field is uniform across the sample.  
It is therefore extremely interesting to ask whether a uniform, 
passively evolving population of galaxies can be found. 

LRGs are the most likely candidate for such a sample, as they are
traditionally assumed to form a single population of galaxies, which
assembled at high-redshift and has been passively evolving since. As
red galaxies dominate the stellar mass budget in the Universe, understanding
how they assemble and grow is a key question in theories of galaxy formation
and evolution. 
Driven both by galaxy evolution and observational cosmology, passive
evolution of LRGs has been tested extensively. Traditionally, this is
done by looking at three main observables and their evolution with
redshift: the total comoving number density, the luminosity function
and density, and the large or small-scale clustering.

Firstly and most simply, passive evolution predicts that the number
density of objects must be conserved as a function of time. This has
been tested by \cite{WakeEtAl06}, who found that the number density of
LRGs brighter than a given luminosity is conserved between $z=0.55$
and $z=0.2$, if one accounts for the passive fading of the stellar
populations.

Secondly, we can go further and look at the evolution of the
luminosity density as a function of redshift. \cite{WakeEtAl06} used
the 2SLAQ sample to investigate whether the luminosity function (LF) at $z=0.2$ and
$z=0.55$ is consistent with a purely passive
evolution. They found that this is indeed the case. Furthermore, they
also looked at the effect of major LRG-LRG mergers in the LF. They
found that the observed LFs are fully consistent with a merger-free
history. \cite{BrownEtAl07}, focusing on the B-band luminosity
density, have found a slight departure from pure passive evolution,
and that around 80\% of the stellar mass in massive red galaxies
($>4L^*$) today was already in place by $z=0.7$. This results in a
modest growth of $\approx 3 \%$ Gyr$^{-1}$ in mass or
luminosity. \cite{CoolEtAl08} put a similar limit on the growth of
very bright red galaxies ($>3L^*$), and showed that bright red
galaxies can not have grown by more than 50\% since $z=0.9$, or
$\approx 6.8 \%$ Gyr $^{-1}$. 

These studies face a fundamental problem that, while we can match
numbers of objects and the luminosity function to those expected for a
particular evolutionary theory, we cannot test whether this is a
coincidence. The match is necessary to test the theory, but is not
sufficient to show that the theory must be true. In particular, for
passive evolution, one can potentially have galaxies entering and/or
leaving the red sequence at any point during the tested timeline, and
several effects can change the number or luminosity density of a
sample. This can happen by:
\begin{enumerate}
\item blue galaxies quenching star formation and becoming red;
\item fainter red galaxies merging together without triggering star
  formation (SF) to enter the sample at lower redshift;
\item LRGs merging with other LRGs to decrease the number density at
  low redshift
\item LRGs leaving the red-sequence due to merger-induced SF.
\end{enumerate}

The lack of massive blue galaxies makes i) and iv) weak
propositions. Furthermore, LRG-LRG mergers tend to be dry, given that
massive red galaxies have little in the way of gas and SF. But ii) and
iii) almost certainly play a role in the overall LRG population
evolution. The question is: how much so?

One can hope to distinguish between these scenarios by investigating
the clustering of the LRGs. The two-point correlation function at very
small scales can be used to infer merger rates, assuming all LRG-LRG
pairs closer than a given distance will eventually merge within a
timescale given by the orbital or dynamical friction time.
\cite{MasjediEtAl06} put an upper-limit on the LRG-LRG merger rate of
$\approx 0.6 \times 10^4$ Gpc$^{4}$ Gyr $^{-1}$ . This very low rate suggests that
mergers between LRGs is not the primary cause of evolution in the
population. In \cite{MasjediEtAl08} the authors extended this
reasoning to other types of galaxies, by computing the
cross-correlation between LRGs and both red and blue galaxies, of
different luminosities. They found that most of the luminosity brought
to LRGs comes from red galaxies via dry mergers and they put an upper
limit on the growth of LRGs at 1.7$\pm$ 0.1 $h$ \% Gyr$^{-1}$ at
$z\approx0.25$. These upper limits sit of the low end of other
measurements, but it should be pointed out the \cite{MasjediEtAl06,
  MasjediEtAl08} measurements only take into account LRG growth that
involves an already existent LRG - i.e., it fails to account for new
LRGs being formed from the mutual merging of fainter (red) galaxies.
Similarly, \cite{deProprisEtAl10} looked at dynamically close pairs in the 2SLAQ sample (concentrating on pairs of galaxies in $0.45<z<0.65$), and found a merger rate of $ 0.6 \times 10^4$ Gpc$^{4}$ Gyr $^{-1}$ , consistent with \cite{MasjediEtAl06} and reaching a similar conclusion.

Finally, one can look at the {\it evolution of the clustering} of the
LRGs and ask whether it is consistent with passive-evolution. At its
simplest, both \cite{WhiteEtAl07} and \cite{WakeEtAl08} found {\it no
  evolution} in the clustering amplitude as a function of
redshift. This alone is a significant problem for passive evolution,
as a scale-independent and deterministic bias predicts evolution in
the clustering strength. To go further, such analyses requires a
framework for modeling this
evolution. \cite{ConroyEtAl07,WhiteEtAl07,BrownEtAl08,WakeEtAl08} have
all, albeit in slightly different fashions, used the halo model to
support their interpretation. Consistently, these studies find that,
within the halo model framework, where there are too many satellites
at low redshift if one is to assume passive evolution from the
observed clustering at high-redshift (33 to 50 per-cent of the
satellites must disappear). These galaxies must either merge with the
central galaxy or be disrupted and become part of the intra-cluster
light (ICL). \cite{ConroyEtAl07} suggest that a significant amount of
the stellar mass in the merged haloes never makes it to the central
galaxy, and that this is a way to reconcile the observed lack of
evolution in the mass or luminosity function of LRGs with clustering
studies that indicate that a large number of mergers happened since
z=1. This argument is given some strength by the fact that they can
match the total stellar mass budget of clusters, and by the prediction
of a significant amount of ICL around satellite galaxies that was
seen in the Virgo cluster \citep{MihosEtAl05}.

Comparison between all of these studies is made very difficult by the
different selection criteria, redshift range, and perhaps most
importantly the number density of the sample, which in turn determines
the luminosity/masses probed. Evolution of galaxies in the red
sequence is heavily dependent on luminosity, with lower-luminosity
galaxies suffering from significant evolution since z=1 (e.g. \citealt{BrownEtAl07}).

\subsection{This work}

In this work we present an analysis of the evolution of LRGs in the
complete sample of galaxies observed using the original SDSS LRG
targeting algorithm \citep{EisensteinEtAl01}. We analyse the number and luminosity density of
these galaxies, the evolution of the luminosity function, and
clustering strength.

Our methodology significantly differs from previous works in two
ways. Firstly, we introduce a new weighting scheme that allows us to
match galaxies at low- and high-redshift whilst keeping most of the
galaxies in the sample. Secondly, as part of our clustering analysis
we analyse a luminosity-weighted power-spectrum and its evolution. On
large-scales, this statistic is insensitive to the merging of galaxies
within the sample assuming conservation of light, and provides
information on the causes of evolution and the population of galaxies
affected. By comparing clustering for luminosity and number-density
weighted samples, we bypass the need to introduce the halo model in 
order to analyse how evolution is occurring.

Throughout the paper, we use the \cite{MarastonEtAl09} fiducial model
for the expected colour evolution of an LRG, assuming passive
evolution (M09). The {\em stellar} evolution is the only model
dependence of this work, apart from dynamical passive evolution that
we aim to test. Our choice of model is further discussed in Sections \ref{sec:ke_corrections} and \ref{sec:model_dependence}.

Our paper is organised as follows: in Section \ref{sec:sample} we
describe our samples, including selection, K+e-corrections,
weighting-schemes and sample matching; in Section \ref{sec:power_spectrum} we explain our method
for the computation of the galaxy power-spectrum and in Section \ref{sec:passive_evolution} its predicted
evolution in the passive scenario; results are presented in Section
\ref{sec:results}; we explicitly consider uncertainties in the stellar model in Section~\ref{sec:model_dependence}; we consider the implications for the LRG
population and present our interpretation in Section \ref{sec:interpretation}; a comparison with previous work is done in Section~\ref{sec:comparison} and finally we summarize in Section~\ref{sec:discussion}.

\section{The LRG sample}\label{sec:sample}

\begin{figure}
\begin{center}
\includegraphics[width=3.5in]{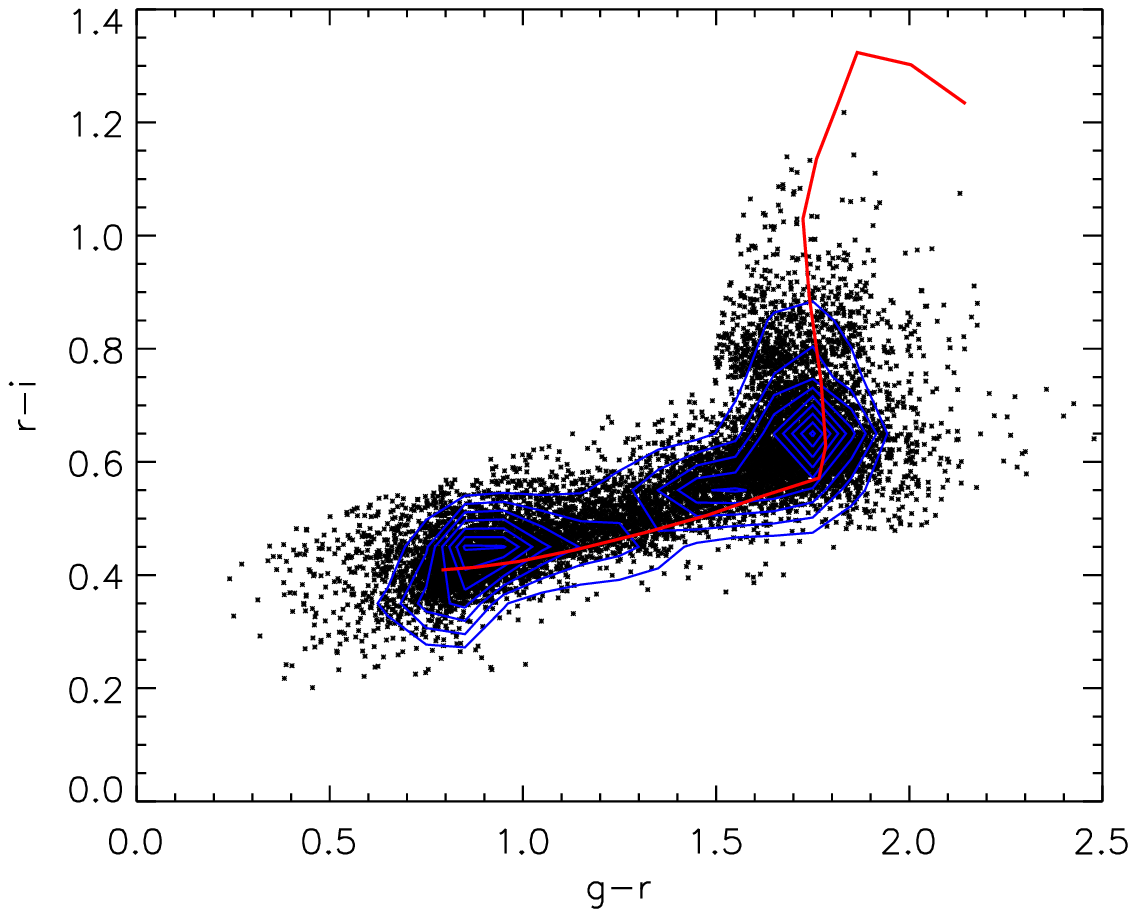}
\includegraphics[width=3in]{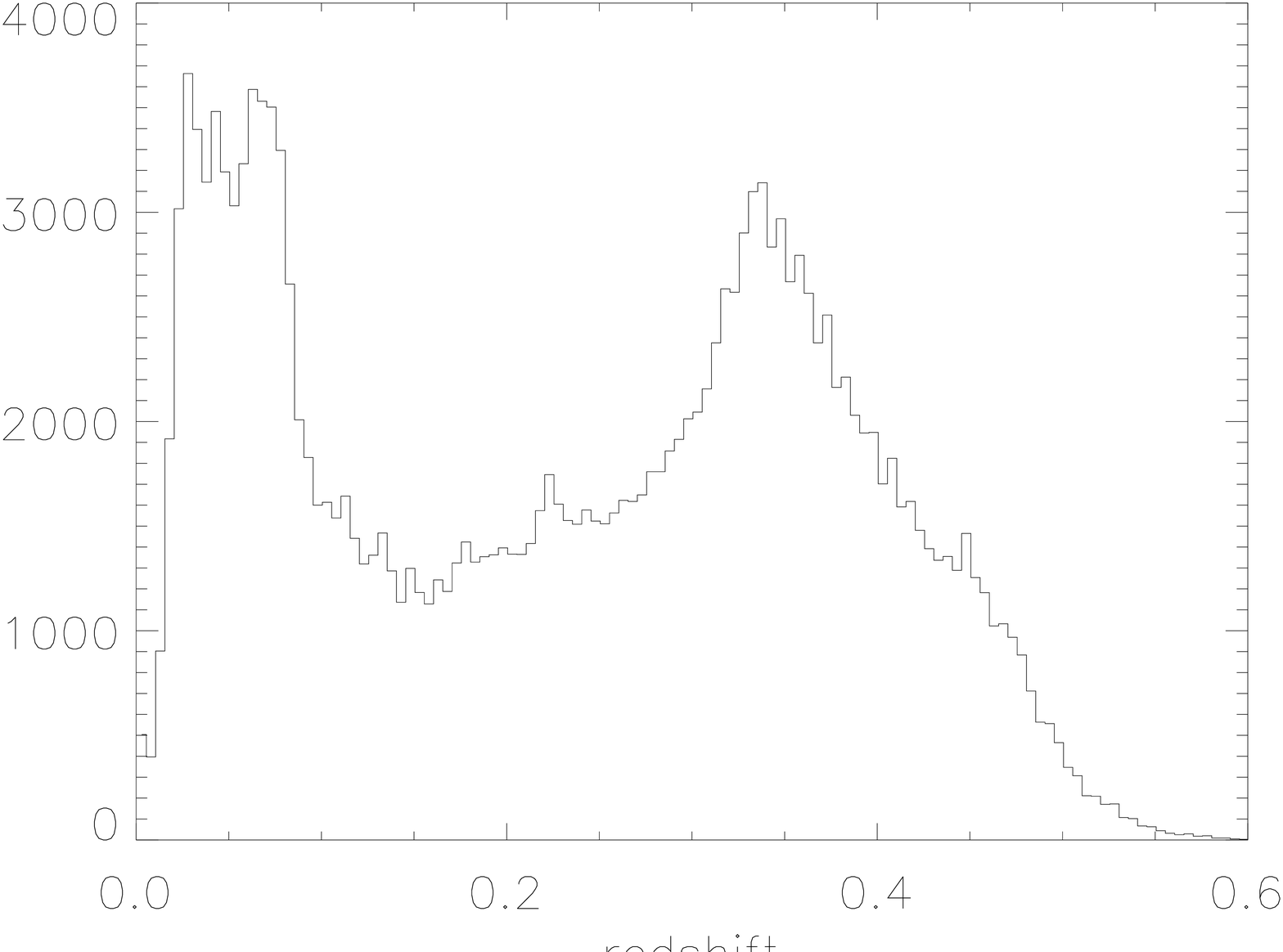}
\caption{{\em Top:} Colour-colour plane for LRGs. The black points are a sub-set of 10,000 galaxies, and the blue contours are representative of the number density of points (as calculated from the full sample). The red line corresponds to the
  fiducial model of Maraston et al (2009). {\em Bottom:} Redshift
  distribution of the original sample. Below a redshift of 0.15 the
  LRG selection is heavily contaminated, and these galaxies are later
  discarded.}
\label{fig:colour-colour-plane}
\end{center}
\end{figure}

We analyse the SDSS spectroscopic LRG sample, whose selection in
colour and luminosity was described in \cite{EisensteinEtAl01}. The
upper panel of Fig.~\ref{fig:colour-colour-plane} shows the
distribution of all of the galaxies in terms of colour. The red line
shows the track predicted by the M09 model for a galaxy made up solely by stars that are 12 Gyr old at redshift of zero, and that have been left to passively evolve since. 
 The lower panel of
Fig.~\ref{fig:colour-colour-plane} shows the redshift distribution of
the galaxies. In this paper we adopt the following nomenclature: SDSS
model magnitudes are written as $u, g, r, i, z$; SDSS petrosian
magnitudes as $u_P, g_P, r_P, i_P, z_P$ and magnitudes given by the
fiducial model of M09 as $u_m, g_m, r_m, r_m, i_m, z_m$.

The M09 model is used to match samples at low and high redshift, and
in calculating rest-frame absolute magnitudes. The SDSS LRG selection,
which was based on the PEGASE models \citep{FiocEtAl99}, is only
important in that this defines the galaxies whose evolution we are
testing. I.e. we can decouple the selection of the galaxies whose
evolution we are testing with the model used to test that
evolution. Essentially, in this paper, we test whether the SDSS LRGs
selected using the \cite{EisensteinEtAl01} cuts are dynamically
consistent with passive evolution, assuming a model for the colour
evolution that we know is able to fit at least part of this sample
\citep{MarastonEtAl09}. One may also want to explore whether samples
of galaxies that are further or closer to the fiducial model, in
colour-colour space, follow passive evolution more or less closely. We
leave that for a follow-up paper.

\subsection{K+e corrections}\label{sec:ke_corrections}

We have used the observed colours and the fiducial model to compute the K- and K+e-corrections used
throughout this paper. The M09 model is the only published model that matches the colour evolution of LRGs over our redshift range of interest, and therefore it is the only viable option for this work. The solution offered in M09 stems from two major additions with respect to the previous literature: the inclusion of a sub-dominant metal-poor stellar component (as opposed to a young stellar component), and a change in the stellar library. \cite{ConroyAndGunn09} have raised some concerns about the latter because they, with their SPS method, could not replicate the changes reported in \cite{MarastonEtAl09}. This inconsistency will be considered further in Maraston et al. (2010, in prep). Additionally, in \cite{MarastonEtAl09}, the authors explicitly assume dynamical passive evolution of the data to which the model is fitted. If this assumption is relaxed one may well find that the inclusion of a metal-poor stellar component is no longer the only - or indeed the best - solution to the problem. The assumption of passive dynamical evolution, however, is the very hypothesis this paper aims to test, and therefore perfectly consistent with our analysis. Thus use of M09 models, which are the best fit to the ridge line of a sub-sample of SDSS selected LRGs, is consistent with the test we are performing.

The fiducial model provides
$L_\lambda(t_{age})$, the luminosity per unit wavelength of an LRG of
age $t_{age}$. To make our analysis more straightforward, we
K+e-correct all galaxies to a common redshift of $z_c=0.1$, and
calculate corrected absolute magnitudes as
\begin{equation}\label{eq:abs_mag}
  M_{r,0.1} = r_p - 5\log_{10} \left\{ \frac{D_L(z_i)}{10 \mathrm{pc}} \right\}
  - Ke(z, 0.1),
\end{equation}
with
\begin{eqnarray} \label{eq:ke_corrs}
  \lefteqn{Ke(z, z_c) =}    \\
  &= -2.5 \log_{10} \left\{ \frac{1}{1+z} \frac{ \int T_{\lambda_o}
      L_{\lambda_o} (z) \lambda_o d\lambda_o \int T_{\lambda/(1+z_c)}
      \lambda_e^{-1} d\lambda_e} {\int T_{\lambda_o/(1+z_c)}
      L_{\lambda_e}(z_c) \lambda_e d\lambda_e \int
      T_{\lambda_o}\lambda_o^{-1} d\lambda_o} \right\}. \nonumber
\end{eqnarray}
$\lambda_o$ is in the observed frame and $\lambda_e$ in the emitted
frame. $T_\lambda$ is the SDSS's $r$-band filter response, and
$L_\lambda(z)$ the luminosity of a galaxy at redshift $z$, given the
fiducial model. Traditionally, calculating K-corrected rest-frame
magnitudes is done to the mean redshift of the sample, and is
accompanied by a change in the filter's wavelength such that, for
galaxies at that redshift (which should be the majority of the
sample), the K-correction is independent of the galaxy's spectrum
(often not known). However, note that Equation \ref{eq:ke_corrs} gives
a fixed K+e correction for a given redshift, and is independent of the
actual observed colours or spectrum of each galaxy. The assumption is
that the model is a good description of all galaxies in the
sample. This makes the choice of the common redshift at which to
normalise spectra, $z_c$, purely arbitrary.

\begin{figure}
\begin{center}
\includegraphics[width=3in]{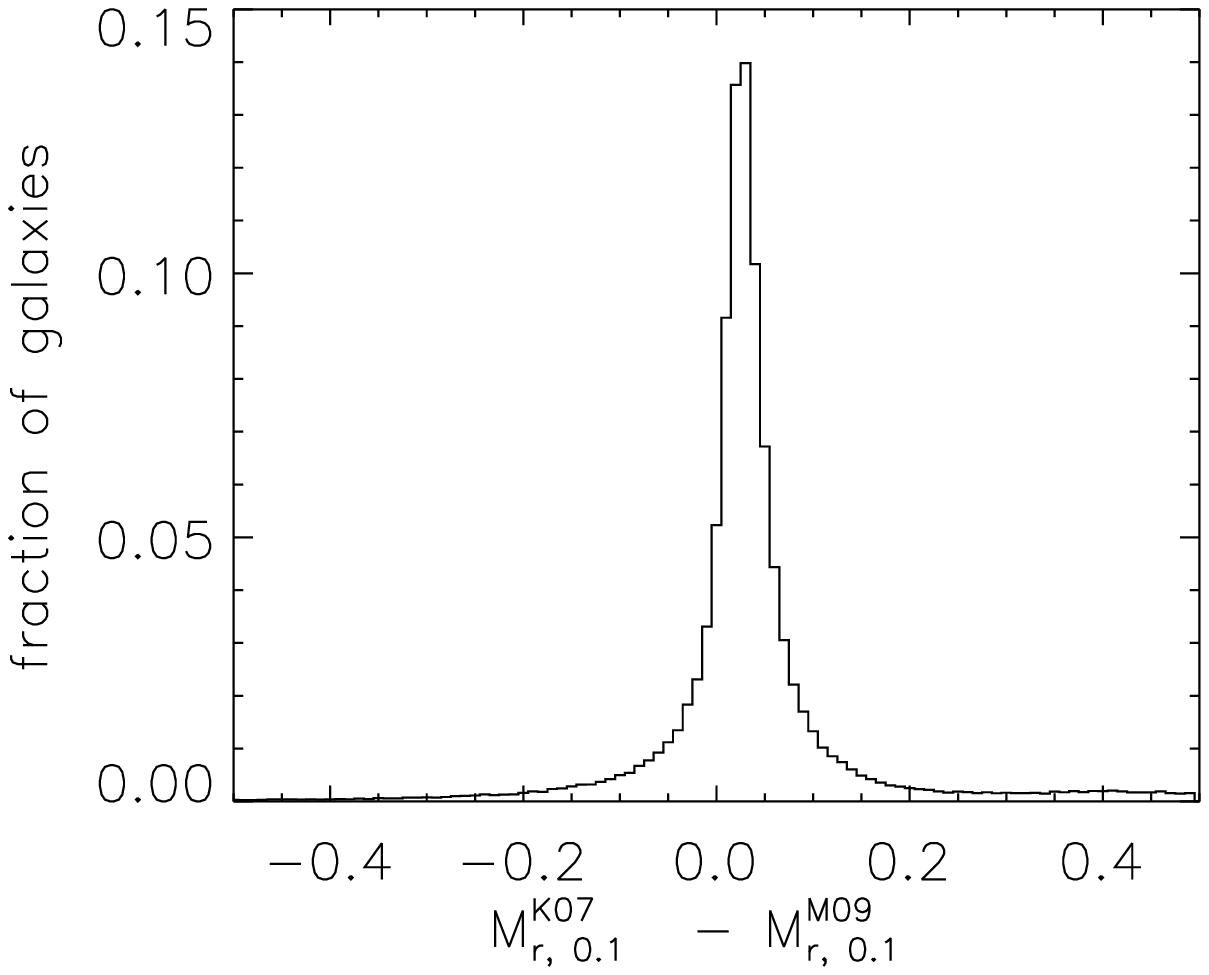}
\includegraphics[width=3in]{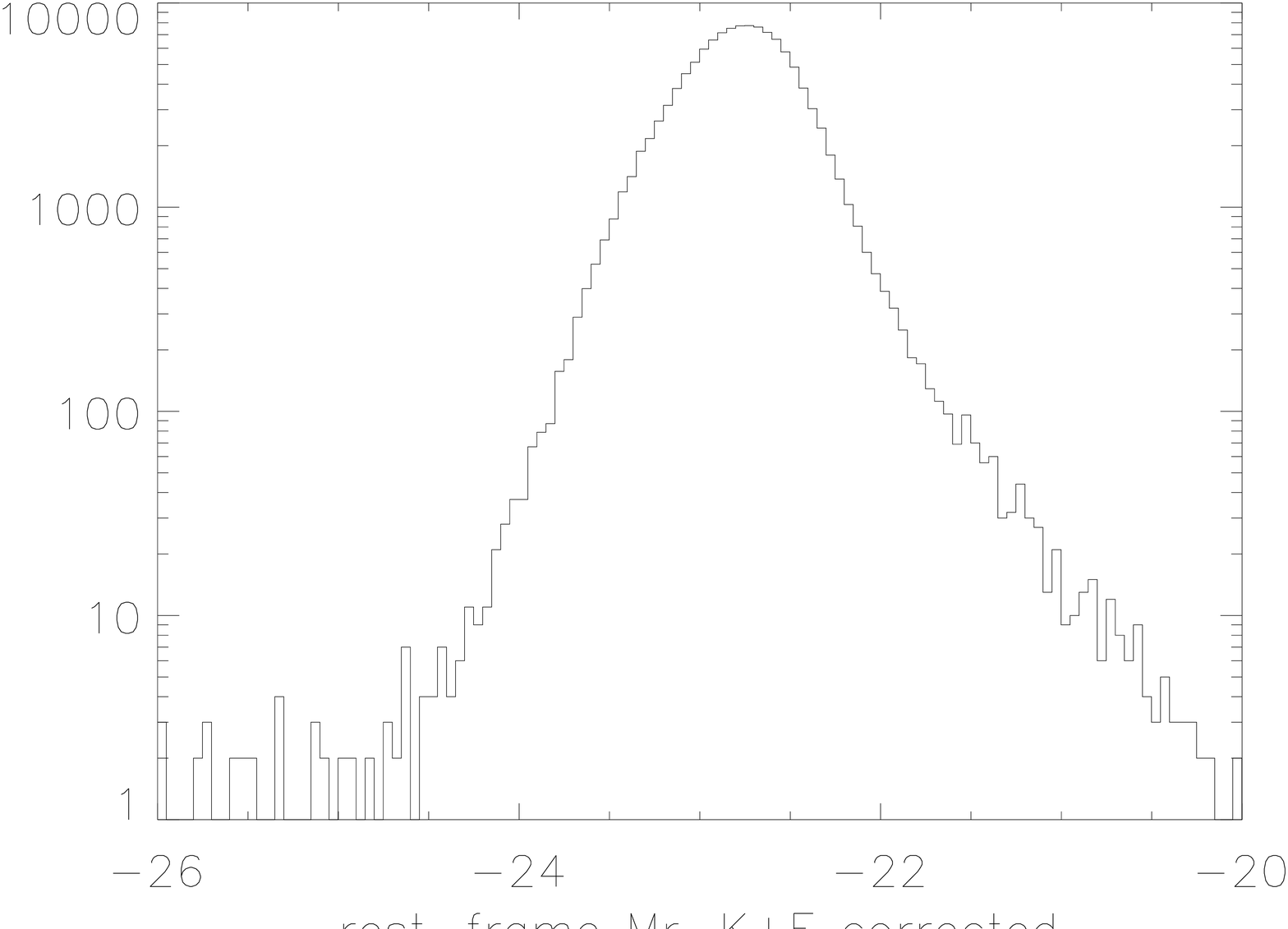}
\caption{{\em Top:} The difference in k-corrected {\it only} rest-frame absolute magnitudes in the $r$-band when obtained with either k-correct (K07) or the M09 model. There is a small offset of around 0.03 magnitudes, which is roughly constant with magnitude.
 {\em Bottom:} Distribution of K+e
  corrected rest-frame magnitudes in the $r$-band for all galaxies at
  $z>0.15$. }
\label{fig:abs_mag_comparison}
\end{center}
\end{figure}

A comparison of the rest-frame $r$-band K-corrected absolute
magnitudes obtained using this model and the ones obtained using the
code K-correct \citep{BlantonEtAl07} can be seen in Fig.~\ref{fig:abs_mag_comparison}. The scatter can be explained by the fact
that K-correct does not use a fixed template, but rather fits a
spectrum to the photometry in order to find the K-corrections. There is a small offset of around 0.03 magnitudes, which is roughly constant with magnitude, and the
agreement is generally good.

\subsection{Sample matching} \label{sec:sample_matching}

In order to analyse the evolution of the LRG population, we construct
a number of pairs of samples, with one sample in each pair at low and
one at high redshift. The primary difficulty in doing this is
selecting a sample at low-redshift that matches, in terms of
individual galaxy properties, the evolved product of the sample at
high-redshift. This sample-matching process has to take into account
four redshift dependent effects:
\begin{enumerate}
  \item the intrinsic evolution of the colour and brightness of an
    LRG,
  \item the varying errors on galaxy colour measurements,
  \item the varying photometric errors in any one band.
  \item the varying survey selection,
\end{enumerate}

Our correction for (i) is unashamedly model-dependent. We work under
the assumption that M09 is a good description of the colour evolution
of the stellar populations, and assume passive dynamical evolution for
the galaxies in our sample - the latter does not pose a problem for
this work since it is the suitability of this model we are trying to
investigate. We include an evolving colour scatter term to allow for
(ii). This is described in Section~\ref{sec:predicting_colours}.

We show that (iii) is a negligible effect in
Section~\ref{sec:sample_selection} by creating mock samples.

Traditionally (iv) has been satisfied by removing galaxies that could
not have been observed in the other sample in the pair
\citep{WakeEtAl06,WakeEtAl08}. However this is wasteful in that,
in order to fully match samples including distributions within each,
we need to remove galaxies that could not have been observed across
all of both samples. In contrast, our approach for (iv) is to
construct a set of weights that assures that each population of
galaxies - in terms of colour and absolute magnitude - is given the
same weight in the high and low redshift samples. For a catalogue
whose selection is only based on a magnitude limit, the traditional
$V_{\rm max}$ estimator could be used for this instead. We explain our
weighting scheme in Section~\ref{sec:weighting_scheme}, which differs
in that it is designed to be optimal in the limit of Poisson errors.

\subsubsection{Predicting LRG colours and magnitudes with redshift}
\label{sec:predicting_colours}

In order to match samples, we need to predict the colour and magnitude
that an observed LRG would have if it was moved to a range of
redshifts following passive evolution. This will depend on i) the
observed colours and absolute magnitude; ii) the fiducial evolutionary
model. In order to match observed samples we also need to account for
the evolution of the colour-colour scatter that has a clear redshift
dependence. 

We are assuming a single evolutionary model, so for any given redshift
the predicted colours are fixed.
To calculate the colours of an LRG at any redshift, we must include
some sort of departure from the fiducial model that takes into
account the observed scatter.

\begin{figure}
\begin{center}
\includegraphics[width=3in]{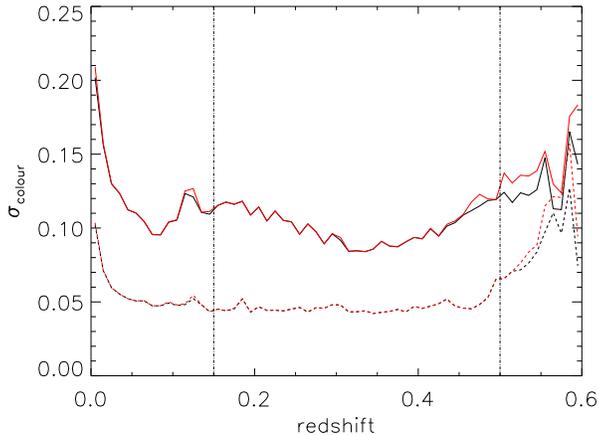}
\caption{$\sigma_{g-r}$ (solid line) and $\sigma_{r-i}$ (dashed line),
  calculated with respect to the fiducial model of Maraston et al
  (2009), as a function of redshift. The red lines are the standard
  deviations obtained after correcting for the fact that we are only
  sampling a finite range in colour and are the ones we use in this
  work. The vertical lines represent the redshift range used. }
\label{fig:color_scatter}
\end{center}
\end{figure}
We start by calculating the scatter with respect to the model, as a
function of redshift, for each of the two colours. This can be seen in
Fig.~\ref{fig:color_scatter}, where we show $\sigma_{g-r}(z)$ and
$\sigma_{r-i}(z)$. Due to target-selection cuts in the colour-colour
plane, we are not guaranteed to fully sample the distribution of $g-r$
and $r-i$, which may affect our estimate of $\sigma_{g-r}(z)$ and
$\sigma_{r-i}(z)$. We are in the regime where this effect is small and we assume that the estimation of the mean is unbiased. We therefore apply a correction to $\sigma$ given by
\begin{equation}
\Delta\sigma^2 = \int_{-\infty}^a (x-\langle x \rangle)^2p(x)dx + \int_b^{-\infty} (x-\langle x \rangle)^2p(x)dx
\end{equation}
where $a$ and $b$ represent the limits sampled by the data, and $p(x)$ is the probability distribution which we assume to be Gaussian. In practice we estimate $a$ and $b$ from the data at each redshift (finding that $|a,b| \gg \sigma$), and use the uncorrected value of $\sigma$ to estimate $p(x)$ and evaluate the integral. Fig.~\ref{fig:color_scatter} shows that the correction is small within the redshift range probed.

Having calculated the scatter, we require that the colour at a
redshift $z'$ of a galaxy observed at $z_0$ departs from the fiducial
model in a way such that
the ratio of the distance from the data point to the model,
$\Delta(g-r)$, and the scatter at that $z_0$ is maintained at any
other redshift. Explicitly:
\begin{equation}
  (g-r)(z') = (g-r)_m(z') 
  + \Delta(g-r)(z_0) \frac{ \sigma_{g-r}(z')}{\sigma_{g-r}(z_0)}, 
\end{equation}
and similarly for $r-i$. Note that we are assuming that the observed scatter in colour is mostly intrinsic or due to photometric errors. I.e., that both cuts are wide enough that they sample the LRG population almost completely. This is not strictly true, but we find that it is indeed almost exactly true, as can be seen by the corrections applied to $\sigma_{g-r}$ and $\sigma_{r-i}$ in Fig.~\ref{fig:color_scatter}.

We calculate petrosian $r$-band magnitudes and surface brightness as a
function of redshift as
\begin{eqnarray}
  r_p(z') &=& M_{r,0.1} + 5\log_{10}\left\{ \frac{D_L(z')}{10 pc} \right\} 
  + \mathrm{Ke}(z', 0.1)\\
  \mu_{50}(z') &=& r_P(z') 
  + 2.5 \log_{10}[2\pi \theta_{50}^2(z')]. \label{eq:sb}
\end{eqnarray}
The K+e-corrections are given by Equation \ref{eq:ke_corrs}. Equation
\ref{eq:sb} assumes that the physical size of a galaxy does not change
with redshift, and $\theta_{50}(z') = \theta_{50}(z_0)
\frac{D_A(z_0)}{D_A(z')}$ with $D_A$ being the angular diameter
distance.

Note that as it is written, the predicted evolution of $r_p(z)$ does
not take into account the changing photometric error with
redshift. The observed scatter in $r_p(z)$ is due both to the increase
in the photometric errors and the scatter in intrinsic luminosity. It
is hard to disentangle the two on a galaxy by galaxy basis, so instead
we test the effect of assuming a constant error with redshift on our
sample matching and selection methodologies. The effect is found to be
negligible, and we give more details in
Section~\ref{sec:sample_selection}.

\begin{figure}
\begin{center}
\includegraphics[width=3in]{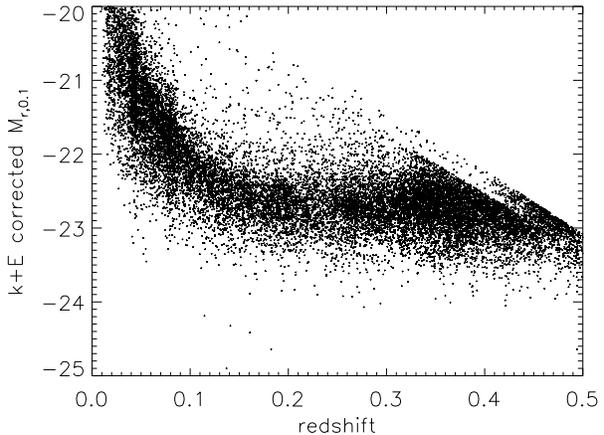}
\caption{K+e corrected absolute magnitude as a function of redshift.}
\label{fig:Mz_full}
\end{center}
\end{figure}
This gives us the colours, petrosian magnitude and surface brightness
as a function of redshift, which allows us to know where each galaxy
would be observed within the survey. Recall that for each galaxy we
also have a redshift, and a K+e corrected $r^{0.1}$-band rest-frame
absolute magnitude. The redshift - absolute magnitude plane can be
seen in Fig.~\ref{fig:Mz_full}. This figure clearly shows that the
SDSS colour cuts result in a sample that doesn't easily lend itself
to a volume-limited analysis. The two sharp diagonal edges correspond
to the two $r_P$-band limits in cut I and cut II at $r_P < 19.2$ and
$r_P < 19.5$ respectively.

\subsubsection{Photometric errors}

As mentioned in Section~\ref{sec:predicting_colours}, there is an added
subtlety in this selection that comes from the fact that photometric
errors increase with redshift. We do not model this in $r_p$ when we
construct and match our samples. That puts us in a potentially
vulnerable position with respect to a Malmquist type of bias, in which
a slope in the number density with luminosity may result in more faint
objects being scattered into a sample than bright objects are
scattered out. Note that if the error was constant with redshift our
weighting scheme would perfectly correct for that because we match the
weight of each population at high and low redshift, for any given
magnitude cut - so we would match any excess or deficit of objects.

To check the effect of assuming a constant error we do the
following. We estimate the true luminosity function of LRGs by using a
\vmax weight on our full catalogue, and use it to construct high and
low-redshift mock catalogues in the magnitude ranges that we adopt for
our final catalogues (see Section~\ref{sec:sample_selection}). We do this
twice for each catalogue: once accounting for the changing error with
redshift in $r_p$, and once assuming a constant error. The effect
in the number and luminosity density in each catalogue between the two
methods is between 0.05\% and 0.3\%, depending on the magnitude cuts. The effect is small enough that we ignore it for the rest of this paper.

\subsubsection{Weighting scheme}\label{sec:weighting_scheme}

In this paper we only consider two redshifts slices, which occupy a
volume $V_A$ (at high-redshift) or $V_B$ (at low-redshift). Using the
analysis described above, we can calculate the maximum volume within
each redshift slice that a galaxy could have been observed in, given
its measured colours and redshift. A traditional $V_{max}$ correction
to match the samples at high and low redshift would use this
information to up-weight galaxies by the reciprocal of the fraction of
the volume of the respective redshift slice within which a galaxy
could be observed. i.e. we should apply a weight to each galaxy
\begin{equation}
  w_i = \frac{V_{\rm slice}}{V_{\rm{max},i}},
\end{equation}
where $V_{max,i}$ is the volume the galaxy could have been observed in
a redshift slice, given the evolution of its optical properties and
the characteristics of the colour cuts, and $V_{slice}$ is the total
volume of the redshift slice the galaxy falls in. Galaxies that can be
observed across all of the slice in which they fall are given a weight
of unity, but galaxies that could only have been observed in part of
the slice are given a larger weight, to compensate for identical
galaxies that exist at other redshifts but fail to meet the survey
selection criteria. Such a weighting scheme fails to remove any
galaxies that are not seen {\em at all} in the other redshift slice,
leaving the match between the two slices unbalanced in terms of galaxy
properties. Furthermore, we also risk massively up-weighting
populations that only exist in very small numbers in a slice, and
Poisson errors of such galaxies may dominate.

Instead, we construct a new scheme that keeps the total weight of each
galaxy population the same in the different redshift slices by {\em
  down-weighting} galaxies based the minimum fractional volume that
the galaxy could have been observed in for both slices. Explicitly,
for a galaxy in $V_A$ we calculate
\begin{equation}\label{eq:wa}
  w_i = \frac{V_A}{V^A_{\rm{max},i}}  \mathrm{min} 
    \left\{ \frac{V^A_{\rm{max},i}}{V_A}, \frac{V^B_{\rm{max},i}}{V_B} \right\},
\end{equation}
and similarly for a galaxy in $V_B$:
\begin{equation}\label{eq:wb}
  w_i = \frac{V_B}{V^B_{\rm{max},i}}  \mathrm{min} 
    \left\{ \frac{V^A_{\rm{max},i}}{V_A}, \frac{V^B_{\rm{max},i}}{V_B} \right\}.
\end{equation}
Where the traditional $V_{\rm max}$ estimator would up-weight
galaxies, we instead down-weight the corresponding galaxies with the
same properties in the other slice.

\begin{figure}
\begin{center}
\includegraphics[width=3.9in]{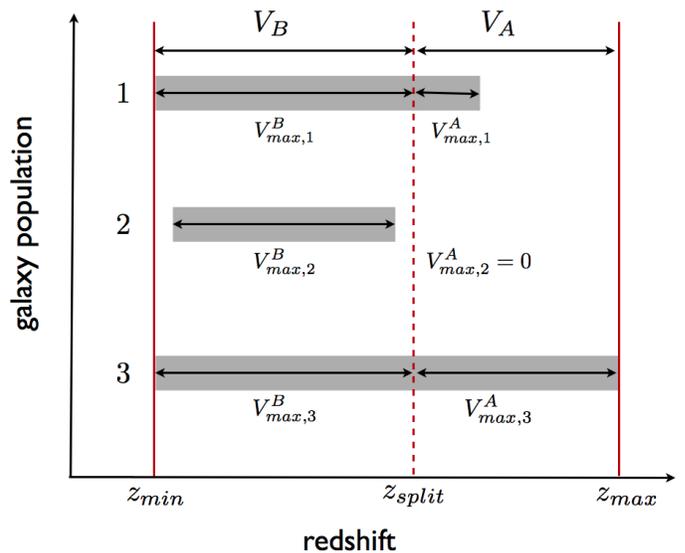}
\caption{A schematic representation of our weighting scheme, given by
  Equations \ref{eq:wa} and \ref{eq:wb}. We give three different
  examples. In each case, the grey stripe represents the volume a
  given type of galaxy could be observed in (see main text for details).}
\label{fig:weighting_diagram}
\end{center}
\end{figure}
In Fig.~\ref{fig:weighting_diagram} we schematically show three
representative populations of galaxies. Here population refers to a
set of galaxies that have similar colours and luminosities. A galaxy
in population 1 that is observed in $V_A$, for example, is given a
weight of unity, according to equation \ref{eq:wa}. A traditional
$V_{max}$ weight would up-weight this galaxy by $V_A/V^A_{max,1}$, but
our approach instead down-weights the population 1 galaxies seen in
$V_B$. This automatically takes care of problematic situations like
population 2, which is constrained to one of the redshift slices - in
this case it would be given a weight of zero. Population 3 simply
represents the galaxies that we can see everywhere in the survey - in
this case the weight is always unity meaning we are of course
volume-limited for this particular population. The weighting is done
on a galaxy-by-galaxy basis.

We have to be careful in the interpretation of results where we use
our weighting, compared with the traditional $V_{max}$
weighting. Whereas the latter gives us the means to correct for
incompleteness and yields true space densities, the former should be
thought as a weighting scheme rather than a completeness
correction. This means that weighted number and luminosity densities
calculated in this way {\em are still potentially volume incomplete},
but the populations are weighted in such a way that they are equally
represented at both redshifts. We can compare the distribution of
total weighted luminosity for the two slices, but we cannot interpret
these functions as giving the luminosity density. 

The advantage of this weighting scheme over simply matching samples by
removing galaxies that do not match the selection criteria in the
other slice is that we keep a large proportion of the sample - in
fact, the number of galaxies thrown out due to the situation of
population 2 in Fig.~\ref{fig:weighting_diagram} is very small. These galaxies contain no information about galaxy evolution between the two samples as galaxies of that type (luminosity, colour) only exist in one of the samples. Consequently, no information is lost by removing these galaxies. It is easy to see that the remaining galaxies are weighted optimally in terms of Poisson statistics: the importance of each sub-population is dependent on the smallest number of galaxies of that type in either sample, and is given a weight accordingly. We
present the final size of our samples, after applying the sample
selection method described in the next Section, in Table~\ref{tab:sample_info}.

\subsection{Sample selection}\label{sec:sample_selection}

In the rest of this paper we consider only the galaxies with redshifts
in $[0.15, 0.5]$. We start by selecting galaxies that are brighter
than a given chosen absolute K+e corrected magnitude,
$M_{r,0.1}^{lim}$, which can be varied. The sample is split at the
median redshift, to ensure that we have the same number of objects in
each slice to begin with. As in Section~\ref{sec:weighting_scheme}, we
have labelled the high-redshift sample as sample~A, occupying a volume
$V_A$ and the low-redshift sample as sample~B, over a volume $V_B$.

We then calculate the integrated {\em weighted} luminosity of the sample as
\begin{equation}\label{eq:Ia}
  I_{A} \equiv \int_{L_{min,A}}^\infty N(L) L  dL = \sum_{ L_i > L_{min,A}} L_i w_i,
\end{equation}
where $N(L)dL$ is the number of LRGs with luminosity in $[L,
L+dL]$. In practice we simply use $L = 10^{-M_r/2.5}$ as a proxy for
luminosity, which is essentially a luminosity in arbitrary units and
$L_{min,A} = 10^{-M_{r,0.i}^{lim,A}/2.5}$. M$_{r,0.i}^{lim,A}$ is
arbitrarily chosen and can be varied so that we may learn about the
evolution of LRGs as a function of magnitude. We can the compute a
weighted luminosity density as
\begin{equation}\label{eq:ellA}
  \ell_A = \frac{I_A}{V_A}.
\end{equation}
This is the luminosity density expected in the $r^{0.1}$-band, at
$z=0.1$, from LRGs in the high-redshift slice, weighted to match the
galaxies observed at low redshift. Alternatively and in an identical
fashion, we can also calculate the weighted number density of the
sample as
\begin{equation} \label{eq:Na}
  N_A \equiv \int _{L_{min,A}}^\infty N(L) dL = \sum_{ L_i > L_{min,A}} w_i,
\end{equation}
and define a total weighted number density as
\begin{equation}\label{eq:nbarA}
  n_A = \frac{N_A}{V_A}.
\end{equation}
We now wish to construct a sample at low redshift that matches either
the luminosity or number density of the sample at high redshift. To do
this we find $L_{min,B}$ such that
\begin{equation}
  \ell_B \equiv  \frac{1}{V_B}\int _{L_{min,B}}^\infty N(L) L dL 
    = \frac{1}{V_B}\sum_{L_i > L_{min,B}} L_i w_i = \ell_A,
\end{equation}
or, alternatively,
\begin{equation}
  n_B \equiv  \frac{1}{V_B} \int _{L_{min,B}}^\infty N(L) dL 
    = \frac{1}{V_B}\sum_{L_i > L_{min,B}} w_i = n_A.
\end{equation}

\begin{table*}
\begin{tabular}{|l|c|c|c|c|c|c|c|c|}
\hline \hline
 & \multirow{2}{*}{$M_{r,0.1}^{lim,A}$} & \multirow{2}{*}{$\bar{z}_A$} & \multirow{2}{*}{$\bar{z}_B$} & \multirow{2}{*}{$z_{split}$} &\multirow{2}{*}{$M_{r,0.1}^{lim,B}$} & \multicolumn{2}{|c|}{Sample size} \\ 
 &							   &                                                   &      					         &      					&   & $V_A$ & $V_B$ \\ \hline \hline
n-matched         & \multirow{2}{*}{-23} & 0.44 & 0.29 & 0.38 & -23.04 & 12,998 & 11,938 \\ 
 $\ell$-matched &                                  & 0.44 & 0.29 & 0.38 & -23.04  & 12,998 & 11,615 \\ \cline{1-8}
 n-matched	 & \multirow{2}{*}{-23.9} & 0.43 & 0.28 &0.37   & -22.95 & 19,567& 16,226 \\
 $\ell$-matched	 &				& 0.43 & 0.28 &0.37	 & -22.96 & 19,567 & 15,650  \\ \cline{1-8}
 n-matched	 & \multirow{2}{*}{-23.8} & 0.43 & 0.28 & 0.36 & -22.89 & 27,328 & 19,433 \\
  $\ell$-matched	 &				& 0.43 & 0.28 & 0.36 & -22.90 & 27,328 & 18,300 \\ \cline{1-8}  
 n-matched         & \multirow{2}{*}{-22.7} & 0.42 & 0.27 & 0.35 & -22.81& 35,632 & 23,654  \\ 
 $\ell$-matched &                                  & 0.42 & 0.27 & 0.35 & -22.83 & 35,632 & 22,135 \\ \cline{1-8}
  n-matched         & \multirow{2}{*}{-22.6} & 0.41 & 0.26 & 0.34 & -22.75 & 43,828 & 27,269 \\
 $\ell$-matched &                                  & 0.41 & 0.26 & 0.34 & -22.77 & 43,828 & 25,209\\ \cline{1-8}
 n-matched         & \multirow{2}{*}{-22.5} & 0.40 & 0.26 & 0.33 & -22.70 & 50,994 & 29,384 \\ 
 $\ell$-matched &                                  & 0.40 & 0.26 & 0.33 & -22.73 & 50,994 & 26,677 \\ \cline{1-8}
 n-matched         & \multirow{2}{*}{-22.4} & 0.40 & 0.26 & 0.33 &-22.68 & 56,632 &  29,687\\
 $\ell$-matched &                                  & 0.40 & 0.26 & 0.33 & -22.72 &56,632 & 26,155\\ \cline{1-8}
 n-matched         & \multirow{2}{*}{-22.3} & 0.39 & 0.25 & 0.32 & -22.65 & 60,828 & 30,855 \\ 
 $\ell$-matched &                                  & 0.39 & 0.25 & 0.32 & -22.70 & 60,828 & 26,915 \\ \hline \hline
 
\end{tabular}
\label{tab:sample_info}
\caption{Basic parameters for the LRG samples analysed in this paper.}
\end{table*}

We can always associate a $M_{r,0.1}^{lim,B}$ with $L_{min,B}$, but
note that it will be different from $M_{r,0.1}^{lim,A}$, even for
samples matched by luminosity-density. This on itself is a potential
estimator of how much a sample deviates from passive evolution.
However, in this paper we will instead concentrate on {\em weighted}
number and luminosity densities alone - a comparison between the
luminosity-density obtained when matching on number-density and
vice-versa may immediately tell us if the data does not support with
passive evolution, we look at this in
Section~\ref{sec:results_densities}. For the rest of this paper, when
we consider the limiting magnitude of the sample we are referring to
the magnitude cut at high-redshift. For reference, we give
$M_{r,0.1}^{lim,A}$ and $M_{r,0.1}^{lim,B}$ for each of our samples in
Table~\ref{tab:sample_info}.

\subsubsection{Choice of $z_A$ and $ z_B$} \label{sec:za_zb}

The entirety of this work is based on comparing two redshift slices. However,
{\it both} the boundary between these two slices and the redshift
distributions within each slice depend heavily on the faint magnitude
chosen. As we include fainter galaxies, we predominantly enrich the sample with low-redshift galaxies, producing a gradient of galaxy density across each sample that will depend on the magnitude limit of that sample. The picture is further complicated as we weight galaxies as described in Section~\ref{sec:weighting_scheme}. We must therefore consider weighted redshift distributions, and use the mean of these weighted
distributions to give a representative redshift for each
slice. Explicitly, 
\begin{equation}
  \bar{z}_A = \frac{ \sum_{z_i \in V_A} {z_i w_i}} {\sum_{z_i \in V_A}  w_i},
\end{equation}
and similarly for $\bar{z}_B$. The values for $\bar{z}_A$, $\bar{z}_B$
and the boundary, $z_{split}$ are given in Table~\ref{tab:sample_info}.

\section{Measuring the clustering}  \label{sec:power_spectrum}

Power-spectra for the galaxy samples were calculated using the method
described by \citet{FeldmanEtAl94}. Each galaxy distribution is converted to an
over-density field by placing the galaxies on a grid and subtracting
an unclustered ``random catalogue'', which matches the galaxy
selection.  To calculate this random catalogue, we model the redshift
distribution of the galaxies using a spline model \citep{PressEtAl92}, and
model the angular mask using a routine based on a {\sc HEALPIX}
\citep{GorskiEtAl05} equal-area pixelization of the sphere as in
\cite{PercivalEtAl09}.

Each galaxy and random is weighted using the method described in
Section~\ref{sec:weighting_scheme}. In order to assign luminosities to
the randoms, we randomly draw a luminosity from the galaxy catalogue,
with the constraint that the redshift of the data point lies within
0.02 of the redshift of the new random point. The weight of the same
galaxy is assigned to the random point - this ensures that the
redshift distributions of the $V_{max}$-weighted samples are the same
for the data and random catalogues. For simplicity, we do not include
a luminosity-dependent bias model that normalizes the fluctuations to
the amplitude of $L_*$ galaxies as advocated by
\cite{PercivalEtAl04}.

Inclusion of the standard \citet{FeldmanEtAl94} weight
\begin{equation}
  w^n_i= \frac{1}{1 + \bar{n}_i \bar{P}},
\end{equation}
would potentially change the match between the high-z and low-z
samples, because the distribution of galaxy number densities in each
slice will be correlated with galaxy properties, and the correlation
may be different for the high-z and low-z samples. We therefore do not
include this weight. Given that the number density of the total SDSS
LRG population is approximately constant with redshift, this only has
a small impact on the error with which we can measure the galaxy
clustering strength.

Power spectra were calculated using a $1024^3$ grid in a series of
cubic boxes. A box of length $4000\mpcoh$ was used initially, but we
then sequentially divide the box length in half and apply periodic
boundary conditions to map galaxies that lie outside the box. For each
box and power spectrum calculation, we include modes that lie between
$1/4$ and $1/2$ the Nyquist frequency (similar to the method described
by \citealt{ColeEtAl05}), and correct for the smoothing effect of the
cloud-in-cell assignment used to locate galaxies on the grid
\citep[e.g.][chap.~5]{HockneyEastwood81}. The power spectrum is then spherically
averaged, leaving an estimate of the ``redshift-space'' power. This
method is the same as that used in \cite{PercivalEtAl09, ReidEtAl09}.

In order to interpret the evolution in the power spectra measured
between high-z and low-z samples, we need to know the effect of the
survey geometry on the recovered power. This window function can be
expressed as a matrix relating the true power spectrum evaluated at
wavenumbers $k_n$, $P_{true}(k_n)$, to the observed power spectrum
$P_{obs,i}$, where $i$ refers to the different band-powers with
central wavenumbers $k_i$:
\begin{equation} \label{defwindow} 
  P_{obs,i} = \sum_nW(k_i,k_n)P_{true}(k_n) - W(k_i,0).
\end{equation}
The term $W(k_i,0)$ arises because we estimate the average galaxy
density from the sample, and is related to the integral constraint in
the correlation function \citep{PercivalEtAl07}. The window
function allows for the mode-coupling induced by the survey geometry.
Window functions for the measured power spectrum (Eqn.~15 of
\citealt{PercivalEtAl04}) were calculated as described in
\citet{PercivalEtAl01}, \citet{ColeEtAl05}, and
\citet{PercivalEtAl07}.

The covariance matrices for the measured power spectrum band-powers
(including correlations between the power spectra from the different
redshift slices), were calculated from $10^4$ Log-Normal (LN)
catalogues \citep{ColesEtAl91,ColeEtAl05}. Catalogues were calculated
on a $(512)^3$ grid with box length $4000\mpcoh$ as in \citet{PercivalEtAl09},
where LN catalogues were similarly used to estimate covariance
matrices. Unlike $N$-body simulations, these mock catalogues do not
model the growth of structure, but instead return a density field with
a log-normal distribution, similar to that seen in the real data. The
window functions for these catalogues were matched to that of the halo
catalogue. The input power spectrum was a $\Lambda$CDM model matched
to the large-scale shape of the observed power spectra.

After removing the effect of the window, we simultaneously fitted the
amplitude of the low- and high-redshift power-spectra amplitudes using
the full covariance matrix. We are interested in the large scales only
- small scales will be affected by intra-halo terms and are sensitive
to merging happening at these scales. The results presented in
Section~\ref{sec:results_clustering} are amplitudes fitted only up to
$k_{\rm max}=0.15$. Our results are robust to changing this scale
cut-off within the range $k_{\rm max} \approx 0.1 - 0.2$. Even though
we use the full covariance matrix when fitting the amplitudes (which
takes into account correlations between the two redshift slices), we
find that the power-spectra are almost independent across the two
slices. The 1-sigma errors shown in
Section~\ref{sec:results_clustering} are therefore simply calculated
assuming this is strictly true.

The \citet{FeldmanEtAl94} methodology assumes that galaxies form a
Poisson sampling of a Gaussian random field. The resulting shot noise,
which arises because of this sampling, is traditionally subtracted
from the measured power spectra. Weighting galaxy populations at
different redshifts by luminosity means that the large-scale
clustering strength is expected to be unchanged by loss-less
mergers. However, the predicted shot-noise would be changed by galaxy
mergers: for example, the expected shot-noise from two equal-weight
galaxies, or one galaxy with twice the weight, are
different. Consequently, for our luminosity weighted samples, we
consider a scenario where we fit both the large-scale amplitude and
the shot-noise, in addition to subtracting the standard Poisson
shot-noise.

\section{Predicting the clustering amplitude} 
\label{sec:passive_evolution}

The observed evolution in the large-scale galaxy power-spectrum is the combination
of the growth of the fluctuations in the underlying density field,
parametrized by the linear growth factor $D(z)$, and the evolution of galaxy
bias set by the peculiar velocity field. We assume that the expected
evolution of the amplitude of the redshift-space power spectrum
between the two redshift slices is given by (\citealt{Kaiser87})
\begin{equation}
  P_g(k, z_B) = \frac{b_B^2+\frac{2}{3}b_Bf_B+\frac{1}{5}f_B^2}
                     {b_A^2+\frac{2}{3}b_Af_A+\frac{1}{5}f_A^2} 
                     \frac{D_B^2}{D_A^2} P_g(k,z_A).
\end{equation}
Here $f$ is the standard logarithmic derivative of the linear growth
factor with respect to the logarithm of the scale factor. For
simplicity, we take $f\simeq\Omega_m^{\gamma(z)}$, $\Omega_m=0.25$ with
$\gamma(z)\simeq0.557-0.02z$, which is accurate to 0.3\% for this
cosmological model \citep{PolarskiEtAl08}.

For a passively evolving population, we assume that the evolution in
the bias is given by \cite{Fry96}
\begin{equation}
  b_z = (b_0 - 1)/D_z + 1,
\end{equation}
where $b_0$ is the bias at zero redshift. In practice, we compute $b_A$ from the data, and calculate 
\begin{equation}
b_B = \frac{(b_A - 1)D_A}{D_B} + 1.
\end{equation}

It is worth noting that the \citet{Fry96} model assumes that the distribution of galaxy velocities matches that of the mass. Simulations show that this is reasonable for the haloes in which LRGs are expected to reside \citep{WhiteEtAl07}, although it is possible to create models, albeit ones that are somewhat contrived, in which this would not hold, such as placing LRGs at stationary points in the density field.


To calculate the growth factor we use the fitting formulae of
\cite{CarrollEtAl92} to calculate the correction to the growth factor from an Einstein-de Sitter to a $\Lambda$CDM model.

\begin{figure*}
  \begin{center}
    \includegraphics[width=2.2in, angle=90]{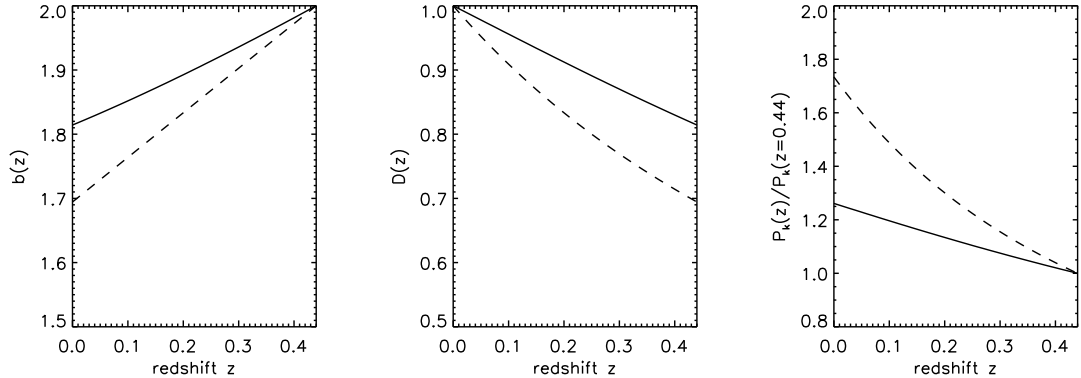}
    \caption{The evolution of the bias (left), growth factor (middle)
      and power-spectrum large-scale amplitude (right) for a
      population of passively evolving galaxies. The dashed line
      assumes an Einstein de-Sitter model, and the solid line includes
      a correction for $\Lambda$CDM.}
    \label{fig:pass_ev}
  \end{center}
\end{figure*}
 Fig.~\ref{fig:pass_ev} shows the expected evolution in the bias,
 growth factor, and overall power-spectrum amplitude for the redshift
 ranges we explore in this paper.

\section{Results} \label{sec:results}

We now present results based on using our high and low redshift
samples to perform the tests of passive evolution described in
Section~\ref{sec:intro} on the number and luminosity densities, the
luminosity function, and the clustering strength.

\subsection{Number and luminosity densities}\label{sec:results_densities}

\begin{table*}
\begin{tabular}{|l|c|c|c|c|c|c|c|c|c|c|}
\hline \hline
\multirow{2}{*}{} & &\multicolumn{2}{|c|}{$n$-matched, \vmatch} & \multicolumn{2}{|c|}{$n$-matched, \vmax} & \multicolumn{2}{|c|}{$\ell$-matched, \vmatch} & \multicolumn{2}{|c|}{$\ell$-matched, \vmax} \\  \cline{3-10} 
                             &  & $n$ & $\ell$ & $n$ & $\ell$ & $n$ & $\ell$ & $n$ & $\ell$ \\ \hline \hline
\multirow{3}{*}{$M_{r,0.1} < -23.0$} & $z_A$ & 4.56e-06 & 8886 & 4.98e-06 & 9705 & 4.56e-06 & 8886 & 4.98e-06 & 9705 \\
						     & $z_B$ & 4.56e-06 & 9082 & 4.76e-06 & 9522 & 4.43e-06 & 8885 & 4.63e-06 & 9308\\
                       				     & {\bf Ratio} & - & {\bf 0.978$\pm$0.0109} & - & - & {\bf 1.026$\pm$0.0133}& - & - & - \\ \hline

\multirow{3}{*}{$M_{r,0.1} < -22.9$}  & $z_A$ &6.44e-06 & 11761 & 7.52e-06 & 13553 & 6.44e-06 & 11761 & 7.52e-06 & 13553 \\
						     & $z_B$ & 6.44e-06 & 12069 & 6.97e-06 & 13016 & 6.24e-06 & 11761 & 6.72e-06 & 12634 \\
                       				     & {\bf Ratio} & - & {\bf 0.974$\pm$0.00949} & - & - & {\bf 1.032$\pm$0.0114}& - & - & - \\ \hline

\multirow{3}{*}{$M_{r,0.1} < -22.8$}  & $z_A$ & 8.38e-06 & 14426 & 1.05e-05 & 17644 & 8.38e-06 & 14426 & 1.05e-05 & 17644\\
						     & $z_B$ & 8.38e-06 & 15023 & 9.36e-06 & 16626 & 7.97e-06 & 14426 & 8.80e-06 & 15822 \\
                       				     & {\bf Ratio} & - & {\bf 0.960$\pm$0.00888} & - & - & {\bf 1.052$\pm$0.0105}& - & - & - \\ \hline

\multirow{3}{*}{$M_{r,0.1} < -22.7$}  & $z_A$ & 1.04e-05 & 17080 & 1.42e-05 & 22423 & 1.04e-05 & 17080 & 1.42e-05 & 22423 \\
						     & $z_B$ & 1.04e-05 & 17770 & 1.24e-05 & 20724 & 9.89e-06 & 17080 & 1.156e-05 & 19626 \\
                       				     & {\bf Ratio} & - & {\bf 0.961$\pm$0.00588} & - & - & {\bf 1.052$\pm$0.00826}& - & - & - \\ \hline

\multirow{3}{*}{$M_{r,0.1} < -22.6$}  & $z_A$ & 1.22e-05 & 19253 & 1.83e-05 & 27287 & 1.22e-05 & 19253 & 1.83e-05 & 27286 \\
						     & $z_B$ & 1.22e-05 & 20102 & 1.54e-05 & 24579 & 1.16e-05 & 19253 & 1.42e-05 & 23033 \\
                       				     & {\bf Ratio} & - & {\bf 0.958$\pm$0.00566} & - & - & {\bf 1.058$\pm$0.00790}& - & - & - \\ \hline

\multirow{3}{*}{$M_{r,0.1} < -22.5$} & $z_A$ & 1.38e-05 & 20975 & 2.23e-05 & 31534 & 1.38e-05 & 20975 & 2.23e-05 & 31534 \\
						     & $z_B$ & 1.378e-05 & 22105 & 1.78e-05 & 27669 & 1.28e-05 & 20975 & 1.61e-05 & 25554 \\
                       				     & {\bf Ratio} & - & {\bf 0.949$\pm$0.00557} & - & - & {\bf 1.072$\pm$0.00783}& - & - & - \\ \hline
				     
\multirow{3}{*}{$M_{r,0.1} < -22.4$} & $z_A$ &1.48e-05 & 21983 & 2.54e-05 & 34543 & 1.48e-05 & 21983 & 2.54e-05 & 34543 \\
						     & $z_B$ &1.48e-05 & 23523 & 1.949e-05 & 29673 & 1.36e-05 & 21983 & 1.70e-05 & 26757\\
                       				     & {\bf Ratio} & - & {\bf 0.934$\pm$0.00557} & - & - & {\bf 1.094$\pm$0.00801}& - & - & - \\ \hline
				     
\multirow{3}{*}{$M_{r,0.1} < -22.3$} & $z_A$ &1.57e-05 & 22811 & 2.81e-05 & 37014 & 1.57e-05 & 22811 & 2.81e-05 & 370134\\
						     & $z_B$ &1.57e-05 & 24478 & 2.10e-05 & 31501 & 1.42e-05 & 22811 & 1.81e-05 & 28113\\
                       				     & {\bf Ratio} & - & {\bf 0.932$\pm$0.00552} & - & - & {\bf 1.099$\pm$0.00798}& - & - & - \\ \hline
\hline
\end{tabular}
\caption{The number and luminosity densities of high ($z_A$) and low ($z_B$) redshift samples for a range of magnitude cuts, for different matching schemes and using different weights. The values shown for a \vmax weight are given only for reference, and all results are based on the \vmatch weights. Magnitudes are K+e corrected to $z=0.1$; number densities are in units of Mpc$^{-3}$ and luminosity densities are in arbitrary units. Note that $z_A$ and $z_B$ are not necessarily the same for the different magnitude cuts. A plot of the ratios in bold in these table is shown in Fig.~\ref{fig:ratios}.}
\label{tab:densities}
\end{table*}

Matching samples based on total numbers of galaxies, obviously allows
us to test how the number density changes between the redshift
slices. If some proportion of the galaxies seen in the high redshift
sample are expected to merge before present day, then the merger
products will be included in the low-redshift sample. In this
situation, matching number density between samples will bring {\em
  extra} galaxies into the low-redshift sample. We should also find
that the total luminosity changes between the samples if no light is
lost to the intra-cluster medium through the mergers. Under the same
assumption, matching samples based on total luminosity will not bring
in these extra galaxies, although we will see a reduction in the total
number of galaxies. Table~\ref{tab:densities} shows our results for a
variety of magnitude ranges; all ratios are of $z_A/z_B$.

\begin{figure}
\begin{center}
\includegraphics[width=3in]{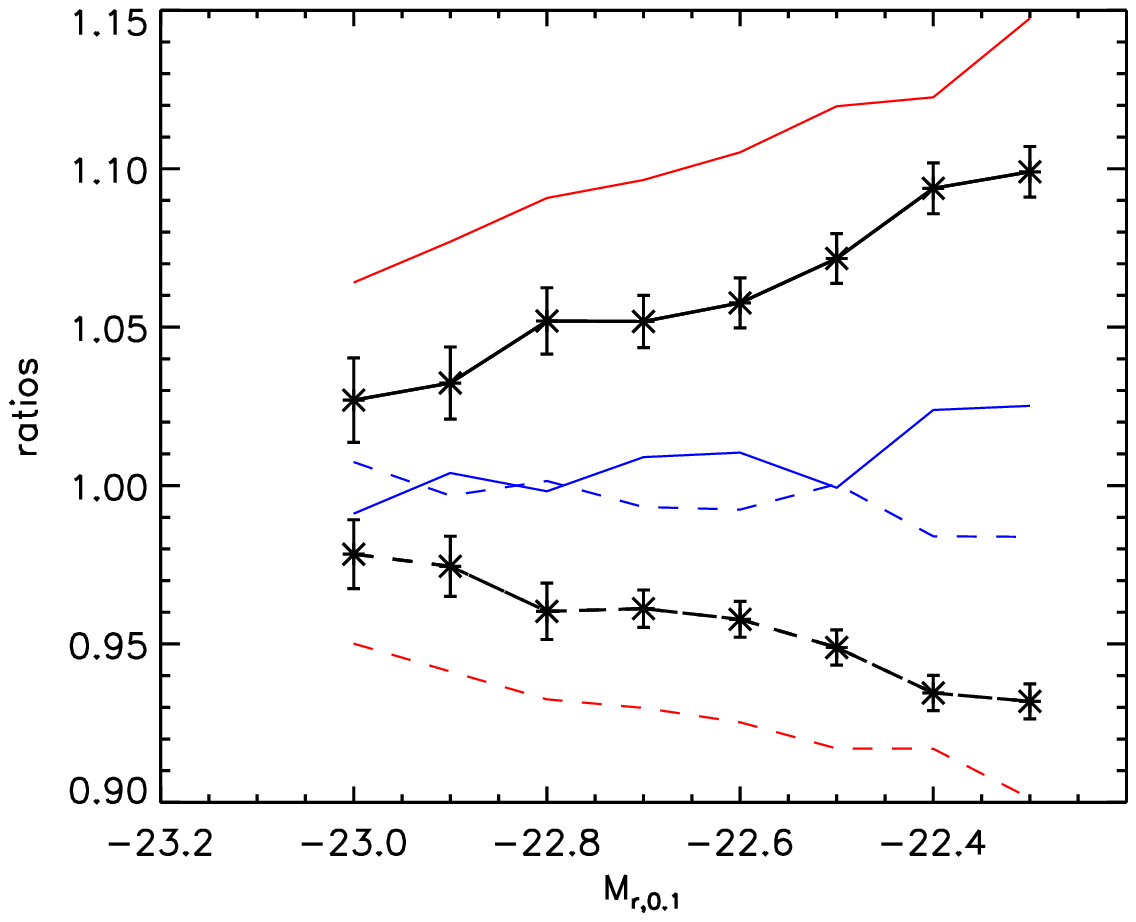}
\caption{The ratio of the number density at high and low redshift when
  matching using the luminosity density (black solid line) and the ratio of
  the luminosity density at high and low redshift when matching using
  the number density (black dashed line). This plot shows the values in bold in Table~\ref{tab:densities}. The blue and red lines show the estimated effect of uncertainties in the IMF slope (see Section~\ref{sec:model_dependence} for details)}
\label{fig:ratios}
\end{center}
\end{figure}
Fig.~\ref{fig:ratios} presents the values in bold in Table~\ref{tab:densities} and shows how the change in number and luminosity
density changes as a function absolute magnitude. The behaviour seen
is perfectly consistent with a scenario where a small proportion of
the LRGs within the sample at high-redshift merge to give brighter
LRGs at low-redshift - we see a decrease in the number density for
luminosity-matched samples and an increase in the luminosity density
for number-density matched samples. This is not, however, the only
explanation. Merging could have also happened between galaxies
outwith the the sample - i.e., galaxies {\em fainter} than the
magnitude cut merging into brighter galaxies between the two
redshifts. As we will see later, our clustering analysis will help us
distinguish between these two scenarios.

It is worth pointing out that the interpretation of Fig.~\ref{fig:ratios} and Table~\ref{tab:densities} is not straight-forward
because the redshift slices change as a function of magnitude. As we
include fainter galaxies in the sample, these are predominantly at low
redshift, and alter the median value that is used to split the
sample. The effect measured is probably a combination of the inclusion
of fainter galaxies and the fact that the split occurs at lower
redshift - the latter is a small effect, but both can act to increase
the number of mergers.

\begin{figure}
\begin{center}
\includegraphics[width=3in]{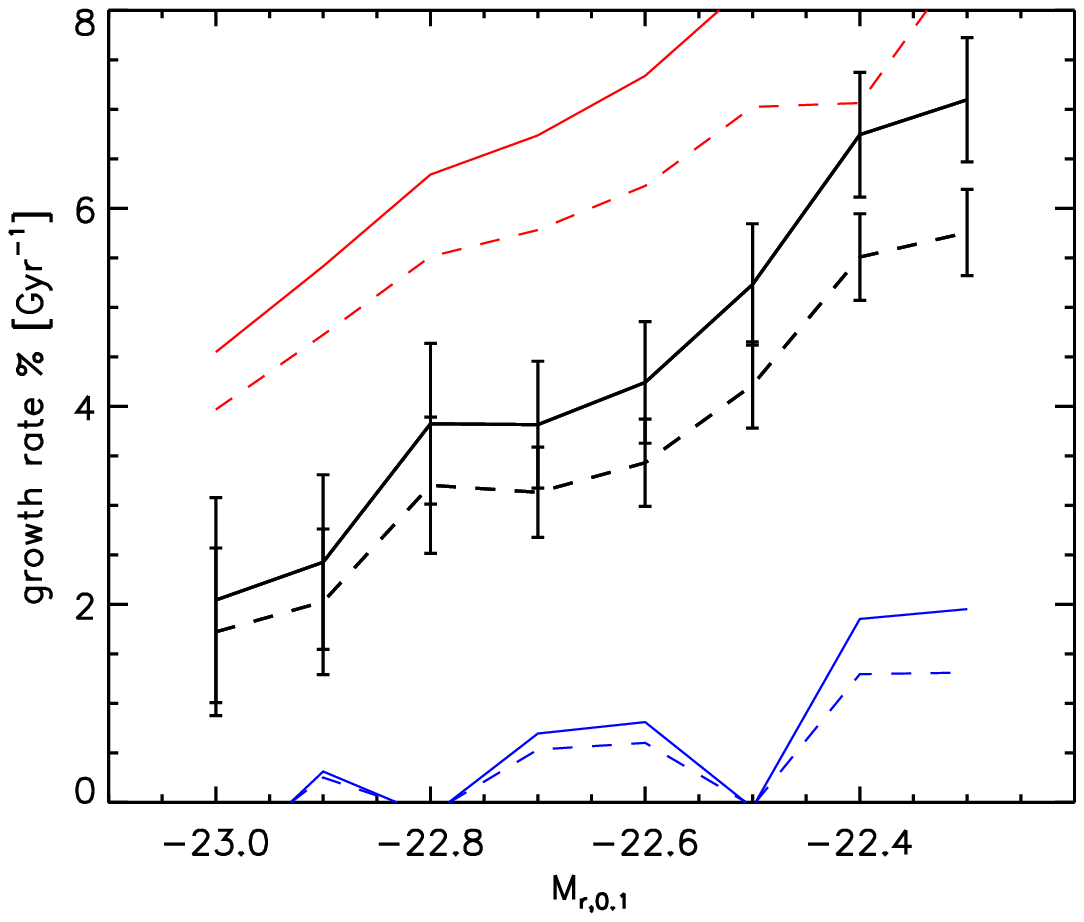}
\caption{The solid black line shows $r_N \times 100$: the percentile growth
  rate in number of galaxies per Gyr, as defined by equation
  \ref{eq:r_N}. This can be interpreted as a merger rate if we assume
  that galaxies in the sample at high redshift have merged. The dashed black
 line shows $r_\ell \times 100$: the percentile increase in
  luminosity density for samples matched to have the same number
  density, as defined by equation \ref{eq:r_ell}. Both quantities are
  shows as a function of the faint magnitude limit, and given in Table~\ref{tab:growth_rates}. The blue and red lines show the estimated effect of uncertainties in the IMF slope (see Section~\ref{sec:model_dependence} for details). }
\label{fig:growth_rates}
\end{center}
\end{figure}
To make the interpretation easier, we can translate the ratio of
galaxy number density into a percentage number of mergers per
Gyr$^{-1}$, following the assumption that differences in number
density are caused by loss-less mergers. This only makes sense in our
luminosity-matched samples, where the low-redshift sample does not
contain extra galaxies. We take the effective time interval $\Delta t$
as being that between $\bar{z}_A$ and $\bar{z}_B$, and calculate
\begin{equation}\label{eq:r_N}
  r_N( M < M_{r,0.1}) = \left(1 - \frac{n_B}{n_A} \right) \frac{1}{\Delta t},
\end{equation}
where $n_A$ and $n_B$ are weighted number densities as defined in
(\ref{eq:Na}) and (\ref{eq:nbarA}). We show this rate, as a percentage
and as a function of magnitude as the solid line in
Fig.~\ref{fig:growth_rates}; the error bars are Poisson errors. Clearly, interpreting this plot as a merger rate
only makes sense if one assumes that LRG-LRG mergers (from within each
sample) are the cause to the change in the number density. If instead
we have fainter galaxies merging together to enter the sample then the
change in numbers can only be interpreted as a more general growth
rate.

We can also do a similar analysis for the fractional luminosity growth
for samples that have been matched to have the same number
density. In this case, we compute
\begin{equation}\label{eq:r_ell}
  r_\ell( M < M_{r,0.1}) 
    = \left( \frac{\ell_B}{\ell_A} -1 \right) \frac{1}{\Delta t},
\end{equation}
where $\ell_A$ and $\ell_B$ are weighted luminosity densities as
defined in (\ref{eq:Ia}) and (\ref{eq:ellA}). We show this fractional
growth rate as the dashed line in Fig.~\ref{fig:growth_rates}. This gain in
luminosity to low redshift, depends on the new galaxies brought into
the low-redshift sample, and cannot therefore be as easily interpreted
by a merger model as the luminosity-matched samples. We quote the values of $r_N$ and $r_\ell$, including Poisson error bars, in Table~\ref{tab:growth_rates}.

\begin{table}
\begin{center}
\begin{tabular}{|c|c|c|}
\hline \hline
$M_{r,0.1}^{lim,A}$ & $r_N \times 100 $ & $r_\ell \times 100$ \\ \hline \hline
-23.0 & 2.04 $\pm $1.035 & 1.72 $\pm $0.847 \\ \hline
-22.9 & 2.43 $\pm $0.883 & 2.02 $\pm $0.735 \\ \hline
-22.8 & 3.82 $\pm $0.812 & 3.20 $\pm $0.689 \\ \hline
-22.7 & 3.81 $\pm $0.640 & 3.13 $\pm $0.455 \\ \hline
-22.6 & 4.24 $\pm $0.614 & 3.43 $\pm $0.441 \\ \hline
-22.5 & 5.23 $\pm $0.612 & 4.21 $\pm $0.436 \\ \hline
-22.4 & 6.74 $\pm $0.630 & 5.51$\pm $0.438 \\ \hline
-22.3 & 7.10 $\pm $0.628 & 5.75 $\pm $0.435 \\ \hline
\hline
\end{tabular}
\end{center}
\caption{The measured values of $r_N$ and $r_\ell$, as defined by equations (\ref{eq:r_N}) and (\ref{eq:r_ell}) and plotted in Fig.\ref{fig:growth_rates}.}
\label{tab:growth_rates}
\end{table}%

\subsection{Luminosity function}\label{sec:results_LF}

\begin{figure}
\begin{center}
\includegraphics[width=3in]{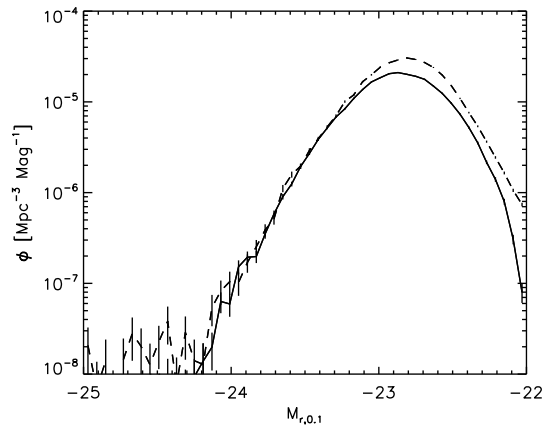}
\caption{The luminosity function in $V_A$ (solid line) and $V_B$
  (dashed line), as defined by Equation \ref{eq:LF}. The redshift
  boundary between the two volumes is $z=0.33$. Given our \vmatch
  weighting scheme, this luminosity function is not corrected for
  volume completeness. Instead, it is constructed such that the
  population at low redshift is perfectly matched to the population at
  high redshift, see text for details. }
\label{fig:LF}
\end{center}
\end{figure}
The next natural step is to construct a luminosity function and study
its evolution, which we compute for each of redshift slices as
\begin{equation} \label{eq:LF}
  \phi (M_{r,0.1}) \Delta M_{r, 0.1} 
    = \sum_{M_i \in  \Delta M_{r,0.1}} \frac{w_i}{V_{slice}}.
\end{equation}
This does not give a luminosity function in the traditional sense, but
rather it gives a population-weighted luminosity function. I.e., the
luminosity function of galaxy samples at two redshifts that represent
the same population of galaxies in equal terms, albeit perhaps
incompletely. We show this pseudo-luminosity function in
Fig.~\ref{fig:LF}.

Fig.~\ref{fig:LF} shows that the information content is not
significantly improved from that in Figs.~\ref{fig:ratios}
\&~\ref{fig:growth_rates}, which show the change in the full sample. This is
simple to understand: small differences in number density are always
diluted when one splits them into magnitude bins. Fig.~\ref{fig:LF}
does make it clear that one needs a way to enrich the low-redshift
sample with galaxies of $M_{r,0.1} > -23$, by means other than
passive stellar evolution.

\subsection{Clustering}\label{sec:results_clustering}

The last test that we perform is based on measuring the evolution of
the clustering of our matched samples of LRGs. The 
\vmatch-weighting makes sure that we are comparing like-with-like
galaxies for clustering measurements. Matching the total luminosity
means that, even allowing for loss-less mergers, we are comparing
galaxies at low-redshift that are evolved products of the
high-redshift galaxies without bringing extra galaxies into the
sample. Additionally, weighting by evolved luminosity, means that the
large-scale clustering should not be affected by these mergers: on
large-scales there is no change in the clustering strength if two
galaxies merge. The method for power spectrum measurement was
described in Section~\ref{sec:power_spectrum}. We compare this
carefully constructed test against the clustering evolution observed
for the samples matched by number density. The passively evolving
model that we test against was described in
Section~\ref{sec:passive_evolution}. 

\begin{figure*}
\begin{center}
\includegraphics[width=2.5in, angle=90]{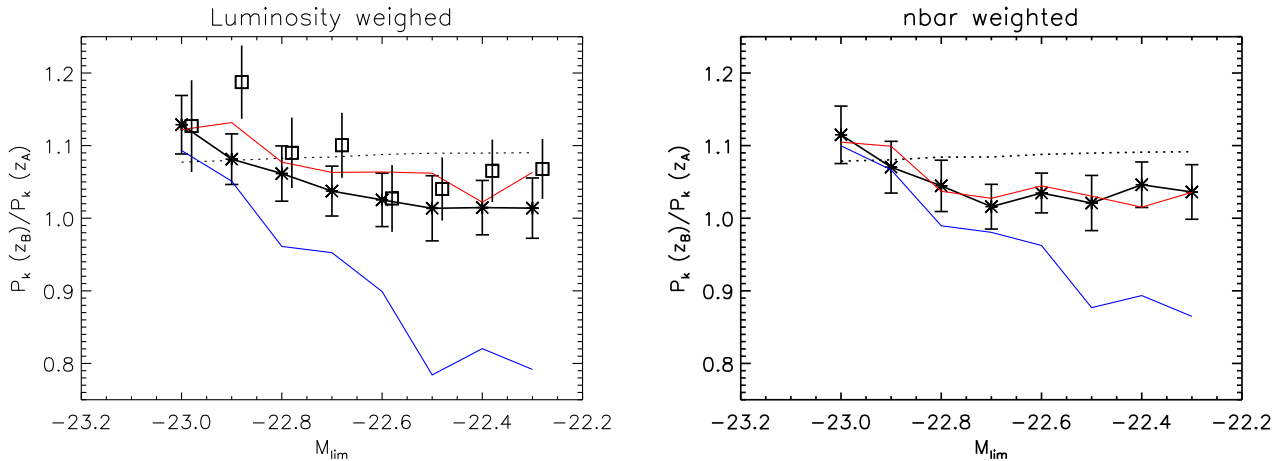}
\caption{The ratio of the large-scale amplitude of the power-spectra
  at high- and low-redshift, as a function of the limiting magnitude
  at high-redshift. Left panel: luminosity-weighted samples and
  power-spectrum. Right panel: number density samples and
  uniformly-weighted power-spectrum. In both panels the dotted line
  shows the expected evolution from a pure passively evolving
  population of galaxies. The open squares in the left panel (slightly offset in magnitude for clarity) show the effect of fitting the shot noise separately, using a $k_{max}=0.2$ - see Section~\ref{sec:power_spectrum} for details. The blue and red lines show the estimated effect of uncertainties in the IMF slope (see Section~\ref{sec:model_dependence} for details). These data points are highly correlated, as
  each fainter sample includes the brighter ones. }
\label{fig:Pk_ratios}
\end{center}
\end{figure*}

We fit an amplitude to the large-scale power-spectra at each redshift,
as described in Section~\ref{sec:power_spectrum}. We are interested in
the evolution of this amplitude from high- to low-redshift, so we will
plot ratios of these amplitudes. Fig.~\ref{fig:Pk_ratios} shows this
ratio as a function of limiting magnitude at high-redshift, for the
luminosity-matched and number density matched samples as the black solid line and stars.

There are four distinct points to appreciate about Fig.~\ref{fig:Pk_ratios}:
\begin{enumerate}
\item neither the luminosity-weighted nor the standard-weighted power-spectrum follow passive evolution at the faint end;
\item in both cases the most luminous objects are found to be consistent with passive evolution;
\item the departure from passive evolution is comparable in both cases; and
\item in both cases, there is an under-evolution of the power-spectrum amplitude with redshift, with respect to that expected from passive evolution.
\end{enumerate}

The departure from passive evolution is less clear in the luminosity-weighted power-spectrum when we fit for shot-noise independently at high- and low-redshift (in the open black squares), which we do for the reasons given in Section~\ref{sec:passive_evolution}. The general behaviour is, however, maintained. A clearer signal is at the moment not possible due to the limited redshift baseline given by our sample.


\section{Model Dependence}\label{sec:model_dependence}

Our results are naturally dependent on the choice of stellar model
that describes the passive evolution of stellar colours and
luminosity. We justify our choice of model by
noting that it describes the {\it colour} evolution of at
least a sub-sample of LRGs (those tested in ~\citealt{MarastonEtAl09})
over a redshift range that goes beyond what is needed
here. Implicitly, we assume that this model is a good description al
{\it all} LRGs in our sample. In this section we explicitly discuss possible ways in which the modelling may be wrong.

The luminosity evolution, for one, is not as well constrained as the colour evolution. We must keep in mind that a model that would predict a
different rate of stellar fading would impact on our
weights that in turn would impact on the sample of galaxies that is
returned by our matching scheme. The dominant factor in this case is the slope of the Initial Mass Function (IMF) at around 1$M_\odot$ \citep{ConroyEtAl09}, but there are in principle other reasons for a change in the luminosity evolution. Although quantitatively motivated by changes to the IMF, the test we present in this section is general for all. 

As explicitly demonstrated in Fig. 8 of \cite{ConroyEtAl09}, a change in the slope of the IMF of around $\pm 0.8$ affects the expected B magnitude change of passively evolving galaxy by as much as 0.4 magnitudes for $z \lesssim 1$. The details of this calculation most certainly depend on the exact modelling of the passively evolving galaxy, as well as on the studied spectral region. However, we can certainly test for an error of this order of magnitude in our samples. We assume that the quoted uncertainty is uniformly distributed since $z=1$ and that $|dM| = 0.2 dz$. We therefore change Equation~\ref{eq:abs_mag} to
\begin{equation}
  M_{r,0.1} = r_p - 5\log_{10} \left\{ \frac{D_L(z_i)}{10 \mathrm{pc}} \right\}
  - Ke(z, 0.1) \pm 0.2z
  \label{eq:dM}
\end{equation}
and re-run our analysis as before. For our redshift slices, which have approximately $dz = 0.15$, the introduced magnitude difference between high and low-redshift will be of the order of $\pm 0.03$ magnitudes with respect to the assumed IMF slope. Note that what is important here is the {\em differential} effect with redshift, which affects galaxies in $V_A$ and $V_B$ with different degrees - we are insensitive to an overall shift in luminosity. In the case of a negative shift ($dM = -0.2dz$), galaxies at lower redshifts will have faded more with respect to galaxies at high-redshift, and we will need {\it more} objects in $V_B$ to explain the luminosity in $V_A$. So whereas before we saw a deficit in the number density when matching the light, we should now find more closely matched number densities. The same reasoning applied to a positive shift ($dM = 0.2 z$), means we should require fewer low redshift galaxies. Note that when we match by number density the main difference comes not from selecting different galaxies, but rather from a different evolution of the magnitudes of these galaxies. The reasoning is similar to the above - for a negative shift we expect the luminosity density at low redshift to be decreased with the respect to the fiducial model. So when we saw an excess in luminosity when matching by number density, we now expect this excess to be reduced. 

We show the effect of a $\pm 0.2z $ change in the fiducial luminosity evolution on the ratios of number and luminosity densities, as well as growth rates, in Figs~\ref{fig:ratios} and \ref{fig:growth_rates}. In each case, the red ($dM = 0.2 z$) and blue ($dM = -0.2 z$) lines show the expected evolution of number and luminosity densities if the luminosity evolution was off by $\pm 0.2z$. One can see that if our description for the luminosity evolution was off by $-0.2z$, then we could just about reconcile the observations with the passive evolution scenario (except perhaps for our faintest sample). In other words, we would be enriching the low-redshift sample with more (lower-luminosity) objects. However, the galaxies removed or included by applying this evolutionary change have the lowest luminosities and are the least biased. This change will consequently also affect the amplitude of the clustering signal.. We show the power-spectrum evolution results in Fig.~\ref{fig:Pk_ratios}. The enrichment of $V_B$ with lower-luminosity objects causes the amplitude of the clustering to be further reduced, as one would expect. Therefore, even though a potential error in the modelling of the luminosity evolution could perhaps explain the evolution of number and luminosity densities without departing from dynamical passive evolution, it could not explain the evolution of the clustering. It is worth commenting on two further aspects of Figs~\ref{fig:ratios}, \ref{fig:growth_rates} and \ref{fig:Pk_ratios}. In all cases, we see a stronger effect in the case of a $-0.2z$ shift, and one which increases to fainter magnitudes. This is probably explained by the slope in the luminosity function at low-redshift. By adding $-0.2z$, we are {\it adding} objects to the low-z sample (whilst keeping the high-z sample more or less the same - see next paragraph), and the results will depend heavily on the luminosity and number densities of these objects (which evolves faster towards fainter magnitudes). The other aspect worth mentioning is that we do not necessarily expect the blue and red lines to bound the solid black line in Fig.~\ref{fig:Pk_ratios}. In this case, the change in the results is given exclusively by the properties of the objects that either enter or leave the sample (with respect to the black line), and given that their number is small then this is effectively a noisy measurement. The exception is given by the increasing offset of the blue line, for which this number is in fact larger for the reasons mentioned above.

There is a final subtlety in the above analysis. When we apply Equation (\ref{eq:dM}), we are effectively changing the absolute magnitude of all the objects in our sample. If we apply the same magnitude cuts as before, we then select a different sample of galaxies and comparison proves difficult. We therefore choose to shift the absolute magnitude limits at high-redshift by the same amount, in order to keep roughly the {\it same objects} in $V_A$ for all cases. Note that both approaches would be valid - in one case we would be studying the effect of Equation (\ref{eq:dM}) for fixed magnitude bins, and in the other we are studying the effect of Equation (\ref{eq:dM}) on a fixed sample of galaxies. 


We also do not directly address the question of contamination. As we
find a clear deviation from passive evolution in at least part of our
sample, we should keep in mind that this could very well be because
the galaxies that reside in the lower-mass haloes are indeed not LRGs
in the traditional, stellar population, sense of the word - i.e.,
their colours may be mimicking those of LRGs due to dust, for
example. This is perfectly OK - we are testing whether the galaxies
that are selected according to the ~\cite{EisensteinEtAl01} target
selection algorithm are passively evolving in a dynamical sense,
independently of what causes their colour evolution. We must only
worry that we have a good model for the description of its colour evolution with redshift.

Finally, let us also consider a scenario in which LRGs are generally passively evolving, but their formation epoch has a dependence on luminosity, or mass. This scenario is easily motivated by the literature, as several authors have found a dependence of mean age with luminosity in early-type galaxies (e.g. \citealt{CaldwellEtAl03, ThomasEtAl05, ClementsEtAl06}), and would effectively mean that the M09 model, even if correct for the most massive LRGs, may become increasingly inadequate for fainter objects. However even the LRGs in our faintest sample, with $M_{r, 0.1} > -22.3$, have typical masses of $10^{11.5} M_\odot$, and velocity dispersions greater than $200$km s$^{-1}$. Given these numbers, we should expect a difference in formation age of less than 1Gyr, which would make little difference for the spectral evolution at the redshifts we are considering. This gives us confidence that the interpretation we present in the next section is driven by the dynamical evolution of the objects in our sample, rather than by an increasingly inadequate model as we go to fainter galaxies.

The dependence on the stellar model is characteristic of all studies
that need to match LRG samples at low and high-redshift, and
emphasises the importance of having a good model for the colour and luminosity 
evolution of LRGs.

\section{Interpretation}\label{sec:interpretation}

In Sections~\ref{sec:results_densities}, \ref{sec:results_LF} and \ref{sec:results_clustering} we have shown the results of testing passive evolution of LRGs using number and luminosity density evolution, as well as the evolution of the clustering of LRGs. If our goal is to simply test dynamical passive evolution of LRGs,
then our three measurements give a very clear answer - LRGs, as
selected, do not {\em all} follow passive
evolution. Fig.~\ref{fig:ratios}, \ref{fig:growth_rates}, \ref{fig:LF}
and \ref{fig:Pk_ratios} consistently show evidence for merging of some
sort. In all cases we also see a clear dependence on luminosity - the
brightest galaxies show the smallest departure from pure passive
evolution. 

To explain the change in weighted number and luminosity densities seen
in Fig.~\ref{fig:ratios} and Fig.~\ref{fig:growth_rates}, we have to
call in on some kind of merging scenario that explains the increasing
luminosity density for samples with the same number density, or the
decrease in the number of galaxies needed to explain the same amount
of light. This can happen in three distinct ways:
\begin{enumerate}
\item merging within the sample;
\item galaxies outside the sample merging with galaxies in the sample; or
\item galaxies outside the sample merging together and getting bright
  enough to make it into the sample.
\end{enumerate}

The clustering result is the least clear, because of the
relatively small redshift range probed. At the same time, however, the introduction of the luminosity-weighted power-spectrum has the most potential to tell us something about the nature of the evolution of LRGs, and differentiate between the three scenarios above.

If we restrict ourselves to the samples matched by number density and
a uniformly-weighted power-spectrum, then our results are in agreement
with previous work from \cite{WhiteEtAl07, WakeEtAl08}, where it was
found that the evolution of the large-scale bias or clustering
amplitude show very little evolution with redshift. A valid
interpretation for this signal is the merging of objects at
high-redshift into one LRG at low-redshift, which decreases the number
density of objects in high-mass haloes and brings the large-scale
clustering amplitude down.

However, if the evolution (or lack of) seen in the right-hand panel of
Fig.~\ref{fig:Pk_ratios} is explained by the merging of satellite LRGs
within the sample, then we expect the luminosity-weighted
power-spectrum to follow passive evolution. The fact that the data in the 
left-hand panel matches that in the right, showing some evidence for a departure from this hypothesis, leads us
to consider a scenario where the lack of evolution in the large-scale power is due
to {\em new} objects entering the sample between the two redshifts.

We can further assert how the new objects are entering the low-redshift sample. If the growth happens at the massive end, we expect the
evolution of the power-spectrum to overshoot that expected from
passive evolution, as more weight is given to objects that are
intrinsically more strongly clustered. In reality, we observe the
opposite behaviour. The only explanation that is consistent with both measurements is
that objects in lower mass haloes are gaining weight from high to low
redshift.

We cannot distinguish between a scenario where galaxies that fall
completely outside the sample at high-redshift come together and
become bright enough to make it into the sample at low redshift, from
a scenario where the faint end of the sample at high redshift is
gaining weight from the merging of companions that initially fall
outwith the sample. These two scenarios, however, are one and the same
- {\it the growth of LRGs, as a population, is happening at the low
  mass end}. As we slide our magnitude limits we probe different
regions of this growth but we observe it increases as we go down in
halo mass/luminosity.

\subsection{Intra-Cluster Light}

We have so far ignored the possibility that light is lost in a merging
event. If this happens then the luminosity-weighted power-spectrum is
no longer expected to follow passive evolution in the case of merging
within the sample. We can ask the question: is there a fraction of
light loss per merger of given mass that can explain the observed lack
of evolution in the luminosity-weighted power-spectrum? To answer this
question quantitatively, we need to know how the amplitude of the
power-spectrum changes as a function of luminosity, $P(k,L)$, which we
leave for a follow-up paper.

Qualitatively however, we can still make the following observation: we
need to increase the weight of objects residing in smaller haloes at
low redshift to explain the observed deficit in power. It follows that any light loss at the
bright end would have to be such that the resulting increase in
luminosity of the bright objects is low enough that the clustering of
these objects does not dominate the overall clustering signal. This
would point towards a differential mass loss fraction with luminosity,
with the most luminous objects losing the most mass to the ICM. The
slope of this relation is related to the slope of $\partial
P(k,L)/\partial L$, but we would still require preferential mass
growth at the low-mass end - in fact, even more so. It therefore seems
inevitable to conclude that the LRG growth is happening predominantly
in lower mass haloes.

\section{Comparison with previous work}\label{sec:comparison}

Our analysis has given us as a behavioural description of the growth
of LRGs. Under our hypothesis, the luminosity growth that we present
in Fig.~\ref{fig:growth_rates} is being introduced by objects that
live in smaller mass haloes, rather than satellite accretion into
luminous central objects. This means that we cannot interpret the
changes in number density between high and low redshift as a merger
rate. The luminosity growth, in turn, is more directly comparable with
previous results.

Even so, the comparison with other work is far from straightforward,
given the different colour/luminosity selection, number density,
redshift range and absolute magnitudes explored in each work. Most
simply, we expect LRG growth to increase with redshift, and with
decreasing luminosity. Table~\ref{tab:literature_summary} presents a
non-exhaustive collection of recent literature results that measured
either the merger rate, or the luminosity growth rate of LRGs using a
variety of techniques. The results that we present in
Fig.~\ref{fig:growth_rates} can at least be said to sample the same
range of values in Table~\ref{tab:literature_summary}, with the
exception of the growth measured by \cite{MasjediEtAl08}. This may be
reconciled with our interpretation if we consider that we need new
objects becoming bright enough to become LRGs form high to low
redshift and that is what dominates LRG growth. By cross-correlation
LRGs with the main galaxy population, the authors concentrate only on
the growth of existing LRGs.

\begin{table*}
\begin{tabular}{|l|l|l|l|l|}
\hline \hline
\multirow{2}{*}{Publication} & \multirow{2}{*}{Redshift range} & Luminosity Growth                          & Merger rate                                & Number density \\ 
                                                   &                                                       & (Gyr$^{-1}$)                                          & (Gyr$^{-1}$)                                 &  (Mpc$^{-3}$) \\ \hline \hline
\cite{MasjediEtAl06}                 & 0.16 - 0.36                                  &                        -                                    & $0.6 \times 10^{4}$ Gpc$^{-3}$              &             -                               \\
\cite{WakeEtAl06}                     & 0.2 - 0.55                                    &                         -                                 & $<10$ \%                                  & $\approx 1-25 \times 10^{-6}$   \\
\cite{BrownEtAl07}                    & $< 0.9$                                      & $\approx 3\%$                                       &                           -                        & $\approx 3 \times 10^{-4}$ \\
\cite{WhiteEtAl07}                     & 0.5-0.7                                        & $< 18 \% $                                             & $\approx 3.4 \%$                       &  $\approx 10^{-3} h^3 $      \\
\cite{CoolEtAl08}                       & 0.1 - 0.9                                      &   $\approx 6.8 \%$                               &                          -                       & $\approx 3-9 \times 10^{-5}$ \\
\cite{MasjediEtAl08}                  & 0.16 - 0.30                                 &   $<1.7 h \%$                           &                  -                                &    -    \\
\cite{WakeEtAl08}                     & 0.2 - 0.55                                    &                          -                                      & 2.4 \%                                        & $\approx 3 \times 10^{-4}$ \\ 
\cite{deProprisEtAl10}	       & 0.45 - 0.65			      &				-			& $0.8 \times 10^{4}$ Gpc$^{-3}$ & 	-			\\ \hline \hline
\end{tabular}
\label{tab:literature_summary}
\caption{A summary of the values obtained in the literature for the growth and merger rates of LRGs. These may be compared with the results we present in Table~\ref{tab:growth_rates} and in Fig.\ref{fig:growth_rates}, but note the varying redshifts and number densities in each study.}
\end{table*}

The luminosity-weighted power-spectrum is a powerful tool to
disentangle merging scenarios: mergers between satellites and centrals
without light loss would give different evolution in the large-scale
clustering strength between number density and luminosity matched
catalogues. Instead we see similar evolution. Additionally, in order
to match the observed decrement in the power spectrum evolution for
the less-luminous galaxies and the increased rate of evolution in
number and luminosity density, the simplest explanation is that LRGs
need to be introduced in smaller haloes which are intrinsically less
clustered. This interpretation is at odds with the one typically
offered by the halo model, which is based on the small-scale
clustering of samples not split by luminosity and that are matched by
number density. In the ``standard'' halo-model based explanation,
departures from passive evolution are due to satellite-central
mergers. The need to match by number density when applying the halo
model to test evolution is on itself a problem in performing these
fits as, if the model suggests some fraction of satellites merge onto
the central galaxy, then the number density at low redshift must also
be reduced. How exactly to do this is not immediately obvious, and has
been done differently by different authors.


\section{Discussion and summary}\label{sec:discussion}

We have, for the first time, applied luminosity weighting to sample
selection and to the power-spectrum, and have presented a new method
with which to interpret the evolution of LRGs. We have also introduced
a new weighting scheme that allows us to keep most of the galaxies in
the sample. This has allowed us to study the evolution of LRGs as a
function of luminosity with unprecedented resolution. We have done
this by measuring evolution of the number and luminosity density of
SDSS LRGs with $0.15 < z < 0.5$, as well as the evolution of their
clustering.

We can summarise our interpretation and conclusions in the following
bullet points:

\begin{itemize}
\item The evolution of LRGs, as a population, is inconsistent with passive evolution.
\item Departure from pure passive evolution is strongest for fainter LRGs; bright LRGs are consistent with pure passive evolution.
\item We see a lack of evolution in the large-scale luminosity-weighted power spectrum for objects with $M_{r,0.1} \lesssim -22.8$, relative to what is expected from passive evolution. This effectively rules out option (i) in Section~\ref{sec:interpretation} as an explanation for the merger and luminosity growth rates presented in Table~\ref{tab:growth_rates}, although the evidence for this interpretation is not strong as a consequence of the relatively narrow redshift range probed.  
\item To explain this lack of evolution instead we propose that {\it LRGs in smaller mass haloes must gain weight (luminosity) since $z=0.5$}. It is unclear whether options (ii) or (iii) of Section~\ref{sec:interpretation} are dominant, but given our sliding magnitude cuts the two are effectively the same process.
\item Our interpretation relies on the assumption that light is conserved when two galaxies merge. However, even if this is not true, any weight given to LRGs in high-mass haloes must be offset by new objects in low-mass haloes entering the sample at $z<0.5$. This, however, only increases the need to introduce less clustered objects into the low-redshift sample. 
\item Our results are not inconsistent with the halo model interpretation, nor previous work, if we restrict ourselves to the same observables. The added information in this paper comes from the matching of the samples on luminosity density (rather than number density) and on the evolution of the large-scale luminosity-weighted power-spectrum.
\item We have explicitly estimated the effect of uncertainties in the IMF slope in the rate of fading of the stellar populations and in the subsequent selection and interpretation of our samples. The departure from passive evolution seen in the number and luminosity densities can certainly be attributed to an (extreme) uncertainty of the IMF slope. However, the resulting clustering signal is then clearly inconsistent with dynamical passive evolution.
\end{itemize}.

In terms of pushing this analysis further, the Baryon Oscillation Spectroscopic Survey (BOSS; \citealt{SchlegelEtAl09}), part of the SDSS-III project, will measure redshifts of LRGs out to $z<0.7$. This will provide the lever-arm required to fully test the evolution of clustering strength as a function of redshift. When combined with luminosity-weighting this will allow us to test whether our interpretation is indeed correct.

\section{Acknowledgments}
This paper benefited from a useful and thoughtful report from our referee, Charlie Conroy. The authors would like to thank Alan Heavens, Martin White, David Wake, Bob Nichol and Claudia Maraston for useful discussions and comments on an earlier draft. RT thanks the UK Science and Technology Facilities Council and the Leverhulme trust for financial support. WJP is
grateful for support from the UK Science and Technology Facilities
Council, the Leverhulme trust and the European Research Council. 

    Funding for the SDSS and SDSS-II has been provided by the Alfred
    P. Sloan Foundation, the Participating Institutions, the National
    Science Foundation, the U.S. Department of Energy, the National
    Aeronautics and Space Administration, the Japanese Monbukagakusho,
    the Max Planck Society, and the Higher Education Funding Council
    for England. The SDSS Web Site is http://www.sdss.org/. The SDSS is managed by the Astrophysical Research Consortium for the Participating Institutions. The Participating Institutions are the American Museum of Natural History, Astrophysical Institute Potsdam, University of Basel, University of Cambridge, Case Western Reserve University, University of Chicago, Drexel University, Fermilab, the Institute for Advanced Study, the Japan Participation Group, Johns Hopkins University, the Joint Institute for Nuclear Astrophysics, the Kavli Institute for Particle Astrophysics and Cosmology, the Korean Scientist Group, the Chinese Academy of Sciences (LAMOST), Los Alamos National Laboratory, the Max-Planck-Institute for Astronomy (MPIA), the Max-Planck-Institute for Astrophysics (MPA), New Mexico State University, Ohio State University, University of Pittsburgh, University of Portsmouth, Princeton University, the United States Naval Observatory, and the University of Washington.

\bibliographystyle{mn2e}
\bibliography{my_bibliography}
\appendix
\end{document}